\documentclass[preprintnumbers,groupedaddress,nofootinbib,aps]{revtex4}
\pdfoutput=1

\usepackage{multirow,array}
\usepackage{amsmath,amssymb}
\usepackage{graphicx}
\usepackage{color}
\usepackage{url}
\usepackage{feynmp}
\usepackage[alsoload=hep]{siunitx}
\usepackage[pdfborder={0 0 0}]{hyperref}
\hypersetup{
  pdfauthor={Dorival Goncalves, Frank Krauss, Silvan Kuttimalai, Philipp Maierhoefer},
  pdftitle={Higgs-Strahlung: Merging the NLO Drell-Yan and Loop-Induced 0+1 jet Multiplicities},
  pdfkeywords={QCD, NLO, Higgs Physics}
}
\begin{document}
\def\contentsname{{\normalsize Content}}
\def\tablename{Table}
\def\figurename{Figure}

\def\bdrs{\text{BDRS}}
\def\pveto{P_\text{veto}}
\def\nj{n_\text{jets}}
\def\meff{m_\text{eff}}
\def\ptmin{p_T^\text{min}}
\def\gtot{\Gamma_\text{tot}}
\def\as{\alpha_s}
\def\az{\alpha_0}
\def\gz{g_0}
\def\w{\vec{w}}
\def\sdag{\Sigma^{\dag}}
\def\s{\Sigma}
\newcommand{\psib}{\overline{\psi}}
\newcommand{\Psib}{\overline{\Psi}}
\newcommand\one{\leavevmode\hbox{\small1\normalsize\kern-.33em1}}
\newcommand{\Mpl}{M_\mathrm{Pl}}
\newcommand{\p}{\partial}
\newcommand{\mat}{\mathcal{M}}
\newcommand{\lag}{\mathcal{L}}
\newcommand{\ord}{\mathcal{O}}
\newcommand{\ope}{\mathcal{O}}
\newcommand{\qqquad}{\qquad \qquad}
\newcommand{\qqqquad}{\qquad \qquad \qquad}

\newcommand{\qb}{\bar{q}}
\newcommand{\matx}{|\mathcal{M}|^2}
\newcommand{\really}{\stackrel{!}{=}}
\newcommand{\msbar}{\overline{\text{MS}}}
\newcommand{\qns}{f_q^\text{NS}}
\newcommand{\lqcd}{\Lambda_\text{QCD}}
\newcommand{\met}{E_T^{\text{miss}}}
\newcommand{\pmiss}{\slashchar{\vec{p}}_T}

\newcommand{\sq}{\tilde{q}}
\newcommand{\go}{\tilde{g}}
\newcommand{\st}[1]{\tilde{t}_{#1}}
\newcommand{\stb}[1]{\tilde{t}_{#1}^*}
\newcommand{\nz}[1]{\tilde{\chi}_{#1}^0}
\newcommand{\cp}[1]{\tilde{\chi}_{#1}^+}
\newcommand{\CP}{CP}

\providecommand{\mg}{m_{\tilde{g}}}
\providecommand{\mst}[1]{m_{\tilde{t}_{#1}}}
\newcommand{\msn}[1]{m_{\tilde{\nu}_{#1}}}
\newcommand{\mch}[1]{m_{\tilde{\chi}^+_{#1}}}
\newcommand{\mne}[1]{m_{\tilde{\chi}^0_{#1}}}
\newcommand{\msb}[1]{m_{\tilde{b}_{#1}}}
\newcommand{\vsm}{\ensuremath{v_{\rm SM}}}

\newcommand{\mev}{{\ensuremath\rm MeV}}
\newcommand{\gev}{{\ensuremath\rm GeV}}
\newcommand{\tev}{{\ensuremath\rm TeV}}
\newcommand{\sign}{{\ensuremath\rm sign}}
\newcommand{\iab}{{\ensuremath\rm ab^{-1}}}
\newcommand{\ifb}{{\ensuremath\rm fb^{-1}}}
\newcommand{\ipb}{{\ensuremath\rm pb^{-1}}}

\def\slashchar#1{\setbox0=\hbox{$#1$}           
   \dimen0=\wd0                                 
   \setbox1=\hbox{/} \dimen1=\wd1               
   \ifdim\dimen0>\dimen1                        
      \rlap{\hbox to \dimen0{\hfil/\hfil}}      
      #1                                        
   \else                                        
      \rlap{\hbox to \dimen1{\hfil$#1$\hfil}}   
      /                                         
   \fi}
\newcommand{\dslash}{\slashchar{\partial}}
\newcommand{\Dslash}{\slashchar{D}}

\newcommand{\eg}{\textsl{e.g.}\;}
\newcommand{\ie}{\textsl{i.e.}\;}
\newcommand{\etal}{\textsl{et al}\;}

\setlength{\floatsep}{0pt}
\setcounter{topnumber}{1}
\setcounter{bottomnumber}{1}
\setcounter{totalnumber}{1}
\renewcommand{\topfraction}{1.0}
\renewcommand{\bottomfraction}{1.0}
\renewcommand{\textfraction}{0.0}
\renewcommand{\thefootnote}{\fnsymbol{footnote}}

\newcommand{\rig}{\rightarrow}
\newcommand{\lrig}{\longrightarrow}
\renewcommand{\d}{{\mathrm{d}}}
\newcommand{\be}{\begin{eqnarray*}}
\newcommand{\ee}{\end{eqnarray*}}
\newcommand{\gl}[1]{(\ref{#1})}
\newcommand{\ta}[2]{ \frac{ {\mathrm{d}} #1 } {{\mathrm{d}} #2}}
\newcommand{\bee}{\begin{eqnarray}}
\newcommand{\eee}{\end{eqnarray}}
\newcommand{\beeq}{\begin{equation}}
\newcommand{\eeeq}{\end{equation}}
\newcommand{\mc}{\mathcal}
\newcommand{\mr}{\mathrm}
\newcommand{\ep}{\varepsilon}
\newcommand{\emt}{$\times 10^{-3}$}
\newcommand{\emfo}{$\times 10^{-4}$}
\newcommand{\emfi}{$\times 10^{-5}$}

\newcommand{\revision}[1]{{\bf{}#1}}

\newcommand{\hzero}{h^0}
\newcommand{\Hzero}{H^0}
\newcommand{\Azero}{A^0}
\newcommand{\PHiggs}{H}
\newcommand{\PW}{W}
\newcommand{\PZ}{Z}

\newcommand{\sw}{\ensuremath{s_w}}
\newcommand{\cw}{\ensuremath{c_w}}
\newcommand{\swd}{\ensuremath{s^2_w}}
\newcommand{\cwd}{\ensuremath{c^2_w}}

\newcommand{\mhhd}{\ensuremath{m^2_{\Hzero}}}
\newcommand{\mhh}{\ensuremath{m_{\Hzero}}}
\newcommand{\mlhd}{\ensuremath{m^2_{\hzero}}}
\newcommand{\Mlh}{\ensuremath{m_{\hzero}}}
\newcommand{\mad}{\ensuremath{m^2_{\Azero}}}
\newcommand{\mhpd}{\ensuremath{m^2_{\PHiggs^{\pm}}}}
\newcommand{\mhp}{\ensuremath{m_{\PHiggs^{\pm}}}}

 \newcommand{\sa}{\ensuremath{\sin\alpha}}
 \newcommand{\ca}{\ensuremath{\cos\alpha}}
 \newcommand{\cad}{\ensuremath{\cos^2\alpha}}
 \newcommand{\sad}{\ensuremath{\sin^2\alpha}}
 \newcommand{\sbd}{\ensuremath{\sin^2\beta}}
 \newcommand{\cbd}{\ensuremath{\cos^2\beta}}
 \newcommand{\cb}{\ensuremath{\cos\beta}}
 \renewcommand{\sb}{\ensuremath{\sin\beta}}
 \newcommand{\tanbd}{\ensuremath{\tan^2\beta}}
 \newcommand{\cotbd}{\ensuremath{\cot^2\beta}}
 \newcommand{\tanb}{\ensuremath{\tan\beta}}
 \newcommand{\tb}{\ensuremath{\tan\beta}}
 \newcommand{\cotb}{\ensuremath{\cot\beta}}



\title{Higgs-Strahlung: Merging the NLO Drell-Yan \\
and Loop-Induced 0+1 jet Multiplicities}

\author{Dorival Gon\c{c}alves,
  Frank Krauss,
  Silvan Kuttimalai,
  Philipp Maierh\"ofer}

\affiliation{Institute for Particle Physics Phenomenology\\
  Physics Department, Durham University\\
  Durham DH1 3LE, United Kingdom}

\begin{abstract}
 \noindent
  We analyse the production of a Higgs boson in association with a Z boson at hadron colliders
  in the Standard Model and some simple extensions. We show how multi-jet merging
  algorithms at leading and next-to-leading order for the loop-induced gluon
  fusion and the Drell-Yan like quark-induced processes, respectively, improve the
  descriptions for various differential distributions, in particular those that
  involve the production of additional jets. The phenomenological studies focus on
  two relevant channels of Higgs boson decays, namely
  $H\rightarrow$~\textit{invisible} and $H\rightarrow b\bar{b}$.  We find sizable
  and phenomenologically relevant corrections to the transverse momentum and
  invariant mass distributions for the Higgs boson candidate.  Thanks to the large
  destructive interference for the top Yukawa terms, this process is very
  sensitive to the magnitude and sign of a possible non-standard top-Higgs
  coupling.  We analyse the impact of this anomalous interaction on distributions
  and estimate constraints from LHC Run II.
\end{abstract}

\maketitle

\section{Introduction}
\label{sec:intro}

The preeminent achievement of Run I of the LHC was the discovery of a scalar
particle resonance~\cite{discovery}, which so far proved largely consistent
with the Standard Model (SM) Higgs boson~\cite{higgs,legacy}.  This discovery
not only marks the end of an era of searches for this elusive resonance, but
it also heralds the beginning of a new era of exploration of the electroweak
symmetry-breaking mechanism.  The increased collision energy and luminosity
of LHC during Run~II allows, in particular, precise measurements of the interactions
of this new resonance with other known particles.  At the same time, other new
resonances interacting with the rest of the SM through the Higgs boson, and new
structures in the interactions of known particles will become a primary ground of
renewed rigorous searches for physics beyond the Standard Model.

In this scenario, the associated production of a Higgs boson with a $Z$ vector
boson, $pp\rightarrow ZH$, also known as {\em Higgs-Strahlung}, is one of the
most prominent paths towards an accurate understanding of the Higgs boson
couplings. Remarkably, this production mode supplemented by jet substructure
techniques can help to access the largest yet most challenging Higgs decay
channel $H\rightarrow b\bar{b}$~\cite{bdrs}, whereas the leading gluon fusion
and vector boson fusion channels fail in this task due to overwhelmingly large
QCD backgrounds.  ATLAS and CMS already have reported first hints for this
process~\cite{zh_atlas,zh_cms}; while the former collaboration provided an
upper limit on the event rate of $1.4$ times the SM expectation, the latter
observed an excess of events above the SM background with $2.1\sigma$.  Run~II
will thus clarify the situation concerning this process, fully establishing
its existence and scrutinising its dynamics.  Ultimately, the $ZH$ channel will
shed light on the highly relevant branching ratio of Higgs bosons decaying
into invisible final states, an important portal for interactions between
the Standard Model and the Dark Matter sector. This channel provides one of the
strongest constraints, where the current upper bounds at 95\% CL reported by ATLAS
and CMS are $\mathcal{BR}(H\rightarrow \textit{inv})<0.75$ and $0.58$~\cite{inv_exp},
respectively.
\bigskip

\begin{figure}[h!]
  $\underset{\mathrm{(a)}}{\includegraphics[width=0.18\textwidth]{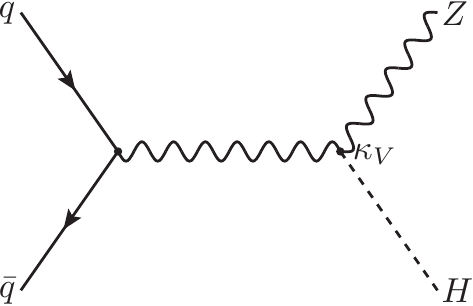}}$\qquad
  $\underset{\mathrm{(b)}}{\includegraphics[width=0.18\textwidth]{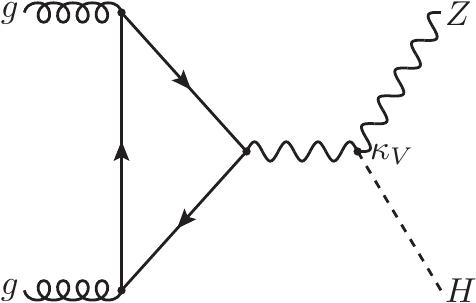}}$\qquad
  $\underset{\mathrm{(c)}}{\includegraphics[width=0.18\textwidth]{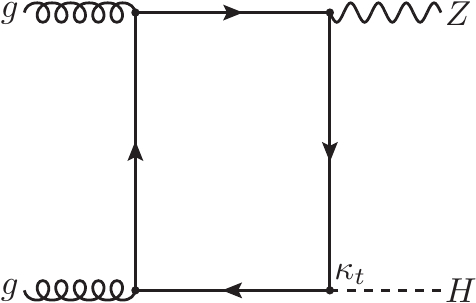}}$
  \caption{Feynman diagrams for $ZH$ production at leading order:
    (a) Drell-Yan-like; (b,c) gluon fusion.
    \label{fig:feynman_diagrams}}
\end{figure}

In the SM, $ZH$ production is dominated by the Drell-Yan-like mode, see
Fig.~\ref{fig:feynman_diagrams}(a).  At leading order (LO), it
contributes to the total cross-section at $\mathcal{O}(\alpha_{EW}^2)$.
Another relevant production mode of $ZH$ final states is gluon fusion,
a loop-induced process mediated by quark loops, depicted in
Figs.~\ref{fig:feynman_diagrams}(b,c) and contributing at LO with
$\mathcal{O}(\alpha_s^2\alpha_{EW}^2)$.  These contributions have been
discussed for example in~\cite{zh_gf}.

At the level of the Feynman diagrams shown, these two sub-processes
do not interfere, but it is important to stress that the latter,
$gg\to ZH$, is part of the next-to-next-to-leading order corrections
to the total $ZH$ production cross section.  Here, we will treat the
two process classes as separate categories, since this allows a more
careful study of the respective QCD emission patterns.  There are
four major factors that guarantee that the gluon fusion process
is larger than the anticipated naive $\alpha_s^2\approx 1\%$:
$i)$ it has a larger initial state colour factor; $ii)$ the process
is driven by the large gluon parton distribution function (PDF);
$iii)$ the top Yukawa coupling $y_t^{SM}$, appearing in the box diagram in the
place of one of the $\alpha_{EW}$ factors, is of order unity
$y_t^{SM}\sim \mathcal{O}(1)$; and $iv)$ the top--quark loop presents
a threshold enhancement at $m_{ZH}\sim2m_t$, which gives rise to
relevant rates at the boosted regime $p_{TH}\sim m_t$. \medskip

On the phenomenological side, and in particular in the framework of
Higgs boson coupling fits, the loop-induced contribution provides an
additional probe to the size and the sign of the top-quark Yukawa
coupling.  In Figs.~\ref{fig:feynman_diagrams}(b,c) the
respective Higgs boson vertices lead to linear terms in $\kappa_t$
and $\kappa_V$, where $\kappa_t$ is a New Physics deviation to the
top-quark Yukawa coupling and $\kappa_V$ represents a potential
rescaling of the $HVV$ interaction vertices
\begin{alignat}{5}
y_t=\kappa_t y_t^{SM},
\hspace{1cm} g_{HVV}=\kappa_Vg_{HVV}^{SM}
\;,
\label{eq:kappas}
\end{alignat}
with $V=Z,W$.  On the other hand, the $WH$ and $qq\rightarrow ZH$
processes probe a single coupling strength, $\kappa_W$ and $\kappa_Z$,
respectively.  At the LHC there are other known processes able to probe
both their size and sign, {\it e.g.}, Higgs boson production in association
with a single top quark $pp\rightarrow tHj$ and the off-shell $H\rightarrow 4$
lepton production~\cite{htj,buschmann2,offshell,legacy}. However, their experimental
observation is challenging due to small rates and huge backgrounds.
\medskip

In many analyses with complex final states, such as the ones emerging
from the $ZH$ processes discussed here, it has become customary to
consider signals and backgrounds in bins of jet multiplicities; one of the
most obvious examples being the process $H\to WW$, where the dominant
$t\bar{t}\to WWb\bar{b}$ background can be fought with jet vetoes.
Similarly, albeit less importantly, the same logic can also be applied
to the $l^+l^- b\bar{b}$ final states typical for $ZH$ production.
There, jet vetoes can play a role similar to considering
boosted topologies, which also suppress the $t\bar{t}$ and similar
backgrounds.  This motivates a more detailed study of jet emission
patterns in this process, where the tool of choice is the multi-jet
merging technology that has already been used in a large number of
Run~I analyses, based on leading-order matrix element calculations.

In this work we study improvements arising from multi-jet merging
techniques applied to the simulation of Higgs-Strahlung, and the impact of the
improvement which these techniques have experienced through the inclusion of
next-to-leading order accurate matrix elements.  The simulation comprises the
following contributions:
\begin{itemize}
\item The Drell-Yan-like ${pp\rightarrow HZ(ll)+0, 1}$~jets at next-to-leading
  order (NLO) accuracy in QCD merged into a single inclusive sample.
  Representative Feynman diagrams are shown below, in
  Fig.~\ref{fig:feynman_diagrams_nlo}.
\item The loop-induced gluon fusion ${pp\rightarrow HZ(ll)+0,1}$~jets
  at leading order merged into a single inclusive sample. Cf.\
  Fig.~\ref{fig:feynman_diagrams_loop2} below for a selection of contributing
  Feynman diagrams and our definition of the 1-jet contribution in this
  channel.
\end{itemize}
Detailed predictions are presented for the invisible
$Z(ll)H(\textit{inv})$ and hadronic  $Z(ll)H(b\bar{b})$ Higgs boson decays.
Using this framework, we also present a realistic phenomenological analysis
deriving anticipated LHC Run II constraints to the $(\kappa_t, \kappa_V)$
coupling parameters. \medskip

This paper is organised as follows. In Section~\ref{sec:zh_sm}, we outline the
basic structures of the Higgs-Strahlung process and point out the impact of
higher jet multiplicities accounted for through multi-jet merging in a large
variety of relevant distributions.  In Section~\ref{sec:yukawa}, we use our
toolkit to explore in detail possible new physics  contributions.  There, we
derive the LHC Run II constraints.  We draw our conclusions in
Section~\ref{sec:summary}.

\section{\protect{$\boldsymbol{ZH}$} production in the Standard Model}
\label{sec:zh_sm}

\begin{figure}[ht]
  \centerline{
    $\underset{\mathrm{(a)}}{\includegraphics[width=0.18\textwidth]{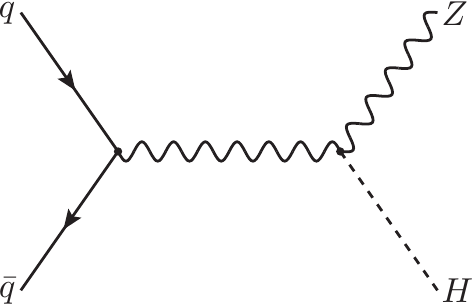}}$\quad
    $\underset{\mathrm{(b)}}{\includegraphics[width=0.18\textwidth]{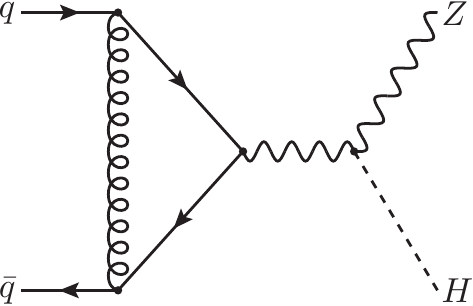}}$}
  \vspace*{2mm}\centerline{
    $\underset{\mathrm{(c)}}{\includegraphics[width=0.18\textwidth]{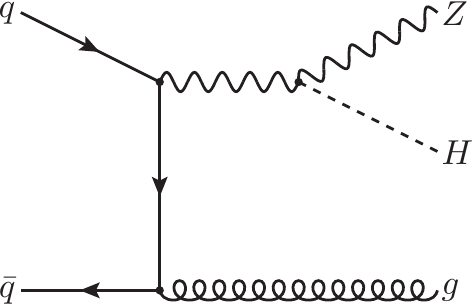}}$\quad
    $\underset{\mathrm{(d)}}{\includegraphics[width=0.18\textwidth]{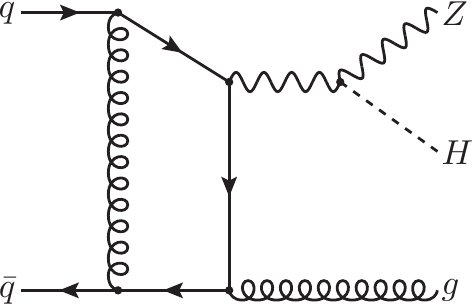}}$\quad
    $\underset{\mathrm{(e)}}{\includegraphics[width=0.18\textwidth]{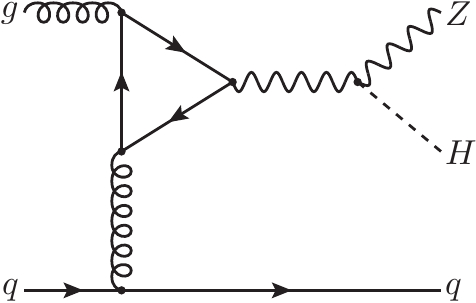}}$\quad
    $\underset{\mathrm{(f)}}{\includegraphics[width=0.18\textwidth]{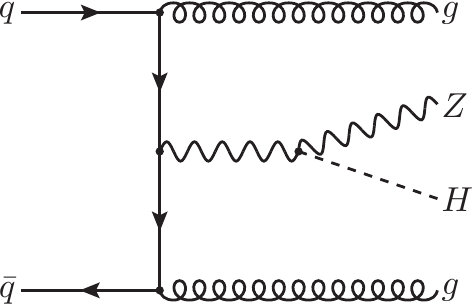}}$}
  \caption{Upper panel: representative Feynman diagrams for Drell-Yan
    $ZH$ production at tree level (a) and at one-loop level (b).
    Lower panel: the same for $ZHj$ at tree level (c), one-loop level (d,e),
    and corresponding real corrections (f).
    \label{fig:feynman_diagrams_nlo}
    \vspace{0.7cm}
    }
\end{figure}
\begin{figure}[h!]
  $\underset{\mathrm{(a)}}{\includegraphics[width=0.18\textwidth]{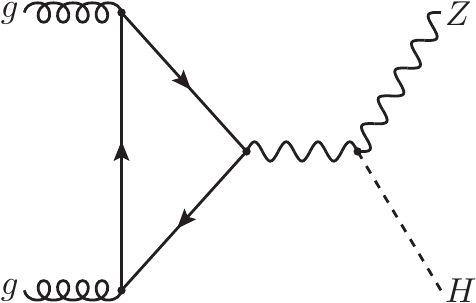}}$\quad
  $\underset{\mathrm{(b)}}{\includegraphics[width=0.18\textwidth]{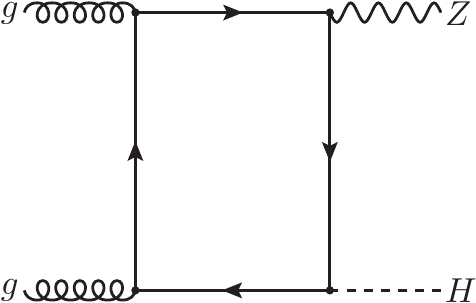}}$\quad
  $\underset{\mathrm{(c)}}{\includegraphics[width=0.18\textwidth]{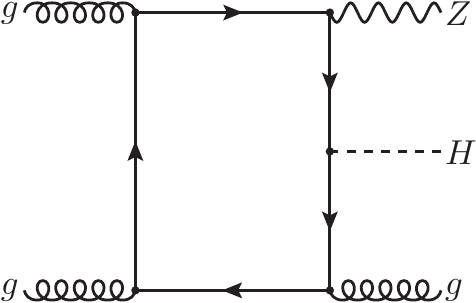}}$\quad
  $\underset{\mathrm{(d)}}{\includegraphics[width=0.18\textwidth]{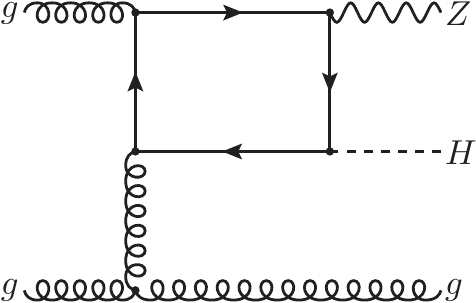}}$\\[2ex]
  $\underset{\mathrm{(e)}}{\includegraphics[width=0.18\textwidth]{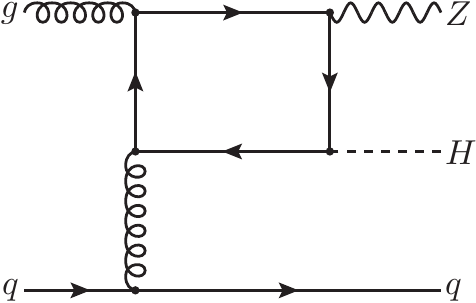}}$\quad
  $\underset{\mathrm{(f)}}{\includegraphics[width=0.18\textwidth]{loopgqhzqTriDY.pdf}}$\quad
  $\underset{\mathrm{(g)}}{\includegraphics[width=0.18\textwidth]{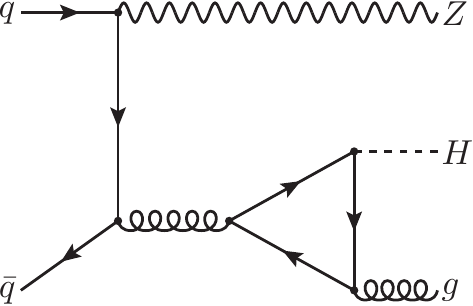}}$\quad
  $\underset{\mathrm{(h)}}{\includegraphics[width=0.18\textwidth]{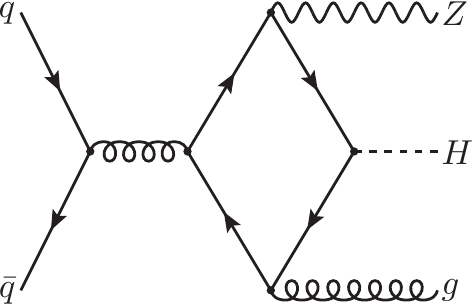}}$
  \caption{Upper panel: representative Feynman diagrams for the Loop$^2$
    contribution. While the gluon fusion contributions to $ZH$ (a,b) and
    $ZHj$ (c,d) are indisputably purely loop induced, the squared loop
    amplitude of diagrams with external quarks and a closed fermion loop
    (e--h) constitute a finite and gauge invariant subset of NNLO corrections
    to $ZHj$. The latter diagrams of course also interfere with the tree level
    amplitude and are therefore included, on the amplitude level, in the
    NLO corrections as well (cf.\ Fig.~\ref{fig:feynman_diagrams_nlo}(e)).
    \label{fig:feynman_diagrams_loop2}}
\end{figure}

\subsection{Higher-order corrections, multi-jet merging, and simulation set-up}
\label{sec:mm}

In this section we discuss in some detail the two dominant $ZH$ production
channels, namely quark-induced Drell-Yan (DY) type Higgs-Strahlung and the
loop-induced gluon fusion (GF), depicted at leading order (LO) in
Fig.~\ref{fig:feynman_diagrams}.  In  particular, we study the impact of
multi-jet merging to NLO accuracy for the DY and to LO accuracy for the GF
contributions.

The DY component comprises the zero- and one-jet squared amplitudes at NLO,
illustrated in Fig.~\ref{fig:feynman_diagrams_nlo} upper and lower panel,
respectively. While the zero-jet GF part
(Fig.~\ref{fig:feynman_diagrams_loop2}(a,b)) is loop-induced and therefore
formally LO, in the context of multi-jet merging it is more convenient to regard
it as a subset of NNLO corrections in the sense of counting coupling constant
powers wrt.\ the DY part. In our definition of GF with an extra jet, besides the
loop-induced diagrams with three external gluons
(Fig.~\ref{fig:feynman_diagrams_loop2}(c,d)), we include all diagrams with a
closed quark loop and an external quark line
(Fig.~\ref{fig:feynman_diagrams_loop2}(e--h)). This definition, like the 0-jet
gluon fusion component, forms a finite and gauge invariant subset of NNLO
corrections to $ZHj$ and captures all diagrams which contain a squared Yukawa
coupling at the squared amplitude level at NNLO QCD. Note that at the amplitude
level there is an overlap of Feynman diagrams between DY and GF. {\it E.g.}, diagram
Fig.~\ref{fig:feynman_diagrams_nlo}(e), interfered with the tree level
amplitude, contributes to NLO DY $ZHj$, while the same diagram
Fig.~\ref{fig:feynman_diagrams_loop2}(g) is also part of the GF amplitude,
contributing at loop-squared NNLO.

\begin{figure}[h!]
  $\underset{\mathrm{(a)}}{\includegraphics[width=0.18\textwidth]{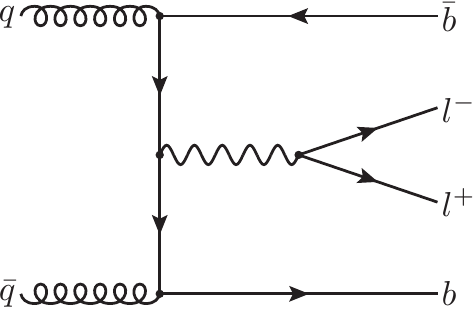}}$\quad
  $\underset{\mathrm{(b)}}{\includegraphics[width=0.18\textwidth]{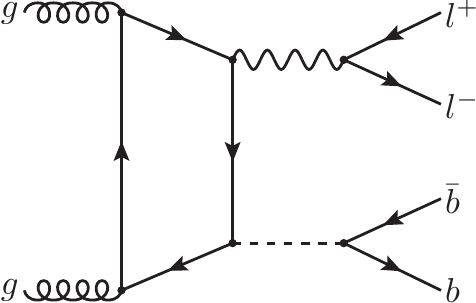}}$\quad
  $\underset{\mathrm{(c)}}{\includegraphics[width=0.18\textwidth]{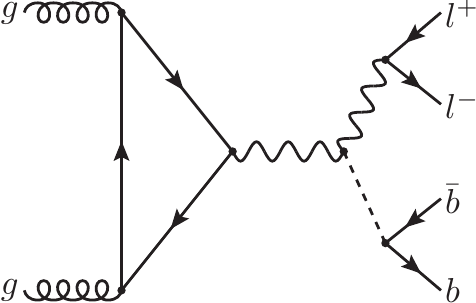}}$\\[2ex]
  $\underset{\mathrm{(d)}}{\includegraphics[width=0.18\textwidth]{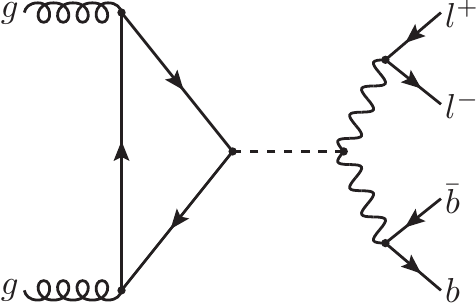}}$\quad
  $\underset{\mathrm{(e)}}{\includegraphics[width=0.18\textwidth]{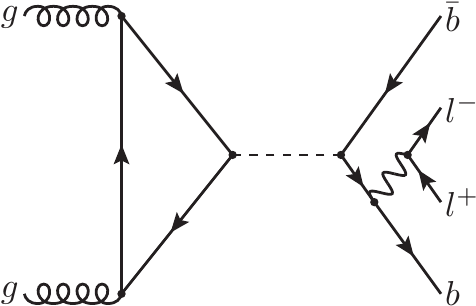}}$\quad
  $\underset{\mathrm{(f)}}{\includegraphics[width=0.18\textwidth]{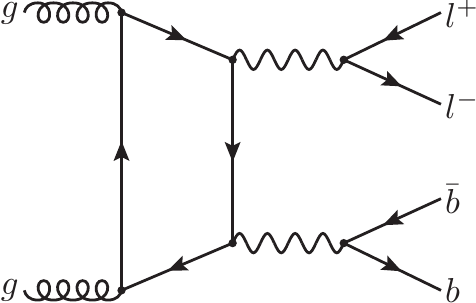}}$\quad
  $\underset{\mathrm{(g)}}{\includegraphics[width=0.18\textwidth]{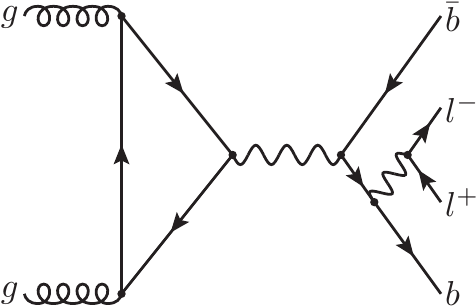}}$\quad
  \caption{Representative Feynman diagrams which contribute to the interference
    of the QCD continuum (a) with the signal (b,c) and background (d--h) when the
    Higgs decays into $b\bar{b}$, or when the $b\bar{b}$ pair is produced via a $Z$
    boson. The loop-induced $ZZ_{GF}$ contribution (f,g) is furthermore included with
    full $l^+l^-b\bar{b}$ final state in order to capture spin correlations and
    off-shell effects.
    \label{fig:feynman_diagrams_interf}}
\end{figure}

Assuming that the invisible sector couples to the Higgs boson only, there is
no interference between signal and background amplitudes in $ZH$,
$H\to inv$. This is not true for $H\to b\bar{b}$ decays, in which case
additional contributions must be considered. Besides the Higgs decay, $b\bar{b}$
pairs can be produced through QCD and through weak interactions, for example via
$Z\to b\bar{b}$. Accordingly, when the $H\to b\bar{b}$ decay is treated as a
part of the matrix elements as shown in
Fig.~\ref{fig:feynman_diagrams_interf}(b,c), the amplitude interferes with the
tree-level QCD continuum $l^+l^-b\bar{b}$ production
Fig.~\ref{fig:feynman_diagrams_interf}(a). Analogously, the tree-loop
interference with diagrams of the kind
Fig.~\ref{fig:feynman_diagrams_interf}(d--g) occurs as a background. In order to
capture spin correlations and off-shell effects in the gluon fusion $ZZ$
background, we take the loop-squared amplitude of diagrams like
Fig.~\ref{fig:feynman_diagrams_interf}(f,g) into account with the full final
state. At this point it is worth mentioning that due to spin considerations
$Z\to b\bar{b}$ and $H\to b\bar{b}$ diagrams do not interfere. Some of these
contributions have not been considered before in the literature.

While multi-jet merged predictions for the DY channel at NLO have been
discussed in~\cite{zh_merging_DY} and the merging in the loop-induced channel
has been technically introduced in~\cite{zh_merging_GF}, here we are mostly
interested in using this technology for detailed studies.  In this context,
it is worth pointing out that the theoretical precision for the DY channel at
fixed order is known up to NNLO in the QCD and up to NLO in the electroweak
perturbative series~\cite{harlander1,nlo_ew}. For the GF contribution, only
estimates~\cite{harlander2} of NLO corrections in the infinite top mass limit
exist; this is due to the fact that a full calculation is hampered by the
presence of many scales and, correspondingly, a prohibitive complexity in the
necessary multi-loop integrals.  Similarly to the gluon fusion process for
Higgs boson and Higgs pair production, the approximation underlying the
estimates mentioned above results in a large correction factor $K\sim 2$ to
the overall cross section.\medskip

The \textsc{Sherpa} event generator~\cite{sherpa} is used throughout this letter, 
supplemented with \textsc{OpenLoops}~\cite{openloops} for the calculation of all loop
contributions and  \textsc{Collier}~\cite{collier} for the evaluation of tensor integrals.  
Finite width effects and spin correlations from the leptonic $Z$ boson decay are fully 
accounted for in the simulation.  For the multi-jet merging at leading order we employ the 
ideas of~\cite{ckkw}, adapted for loop-induced processes in~\cite{WW_+_jets}.  For the 
merging of next-to-leading order matrix elements, the \textsc{MEPS@NLO}
algorithm~\cite{mepsnlo} is used. This is implemented along the
standard LO multi-jet merging algorithms in \textsc{Sherpa}, which also
provides tree-level amplitudes, tools for infrared subtraction in
the calculation of NLO QCD cross sections, and the simulation of parton
showers, hadronisation, hadron decays, etc.~\cite{sherpastuff}.  Throughout
our studies we use the \textsc{NNPDF3.0} parton distribution functions at
NLO accuracy~\cite{nnpdf}.  In our estimates of theoretical
uncertainties, we focus on the usual renormalisation and factorisation scale
variations by up to a factor of two  in the fixed-order part of the simulation. 
From all these possible scale choices we omit the ones in which the factorisation 
and renormalisation scale prefactors differ by a factor of $4$.  Since there is no 
higher-order calculation for the GF contribution, we not only consider the usual scale
uncertainties, but we also consider the effect of higher-orders on the
total cross section ({\it i.e.}, the GF rates account for $K=2$).
Since the $K$-factor estimate most likely is too naive
an assumption, we estimate the associated uncertainty by varying the
$K$-factor in the range from $1.0$ to $4.0$.  We compare the
resulting error bands with the ones obtained from the customary scale
variations in Section~\ref{sec:inv}.  In addition, there are two more sources
of uncertainties stemming from the combination of the fixed-order matrix
elements with the parton shower.  The first one is related to the jet cut
used in the multi-jet merging, $Q_\text{cut}$, which we vary according to
$Q_\text{cut}/\si{\giga\electronvolt}\in\{15,20,25\}$.
In addition, the uncertainty related to the resummation performed numerically
in the parton shower is estimated by varying the starting scales with factors
in $\{\sqrt{0.5},1.0,\sqrt{2.0}\}$.
\medskip

\subsection{Invisible decays: $\boldsymbol{Z(ll)H(\textit{inv})}$}\label{sec:inv}

Invisible Higgs decays occur in many models collectively referred to as
``Higgs-portal'' models, see {\it e.g.}~\cite{hportal}.  In these models,
the Higgs boson is the mediator between the SM particles and an unknown
sector with no other tangible interactions with the other Standard Model
fields and therefore a prime candidate for Dark Matter.  Higgs boson
production in association with a $Z$ is a particularly suitable channel for
invisible Higgs decay searches due to its clean signature with large amounts
of missing energy from the undetectable Higgs decay products recoiling against
the (boosted) leptonic $Z$ boson decays.\medskip

\begin{figure}[t!]
 \includegraphics[width=.55\textwidth]{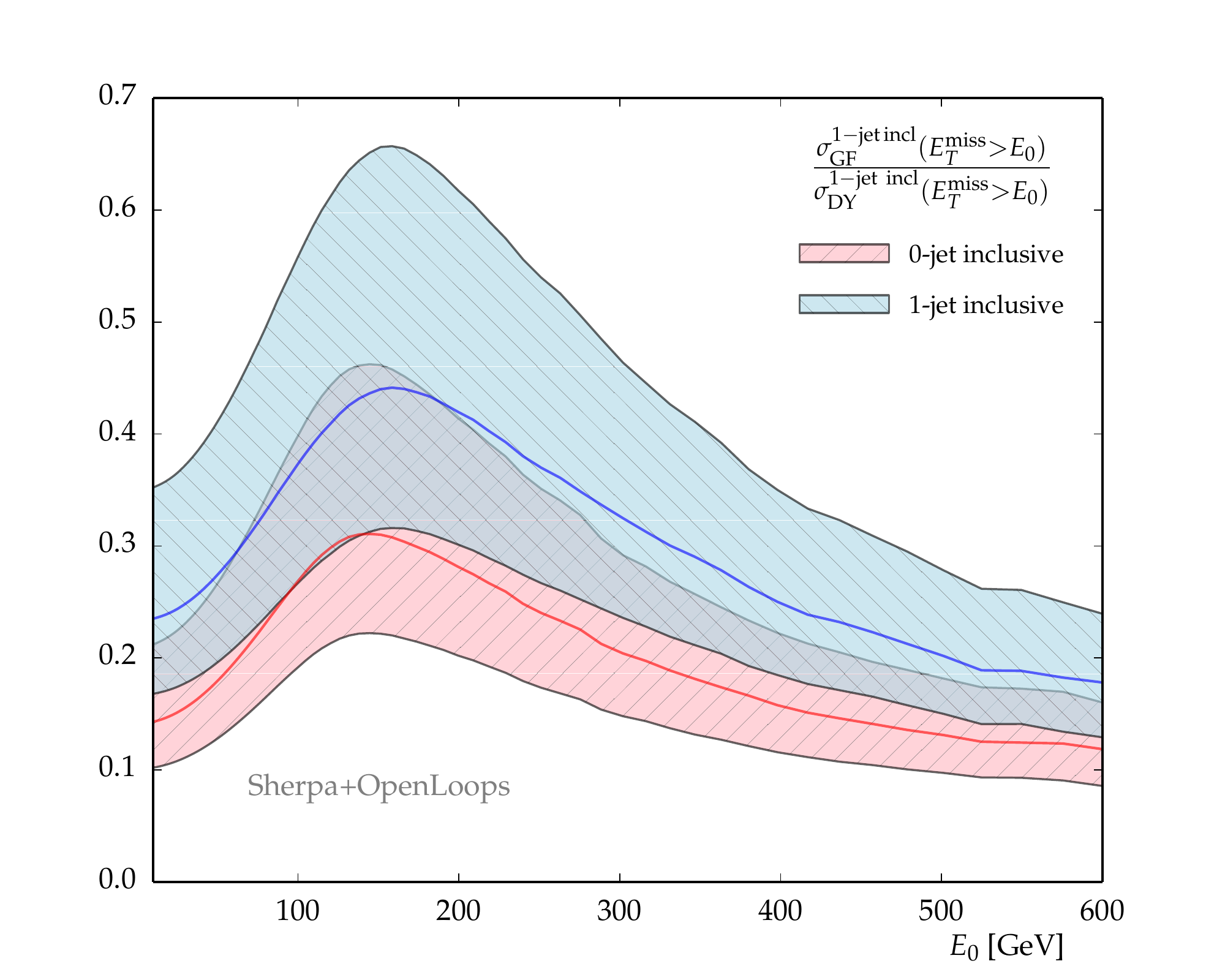}
 \caption{Relative size of the gluon-fusion contribution to the cross
   section as a function of a minimum $E_T^\mathrm{miss}$ cut. We show
   individual curves for the total inclusive cross section and the
   1-jet inclusive cross section. Uncertainty bands are obtained from
   scale variations in the gluon fusion sample, keeping the
   denominator fixed. The NLO Drell-Yan and the
   loop-induced gluon fusion samples are both
   merged up to one jet, respectively denoted as \textsc{MEPS@NLO} and \textsc{MEPS@Loop$^2$}}
  \label{fig:xs_minmet}
\end{figure}

\begin{figure}[ht!]
  \hspace{-0.2cm}
  \includegraphics[width=.5\textwidth]{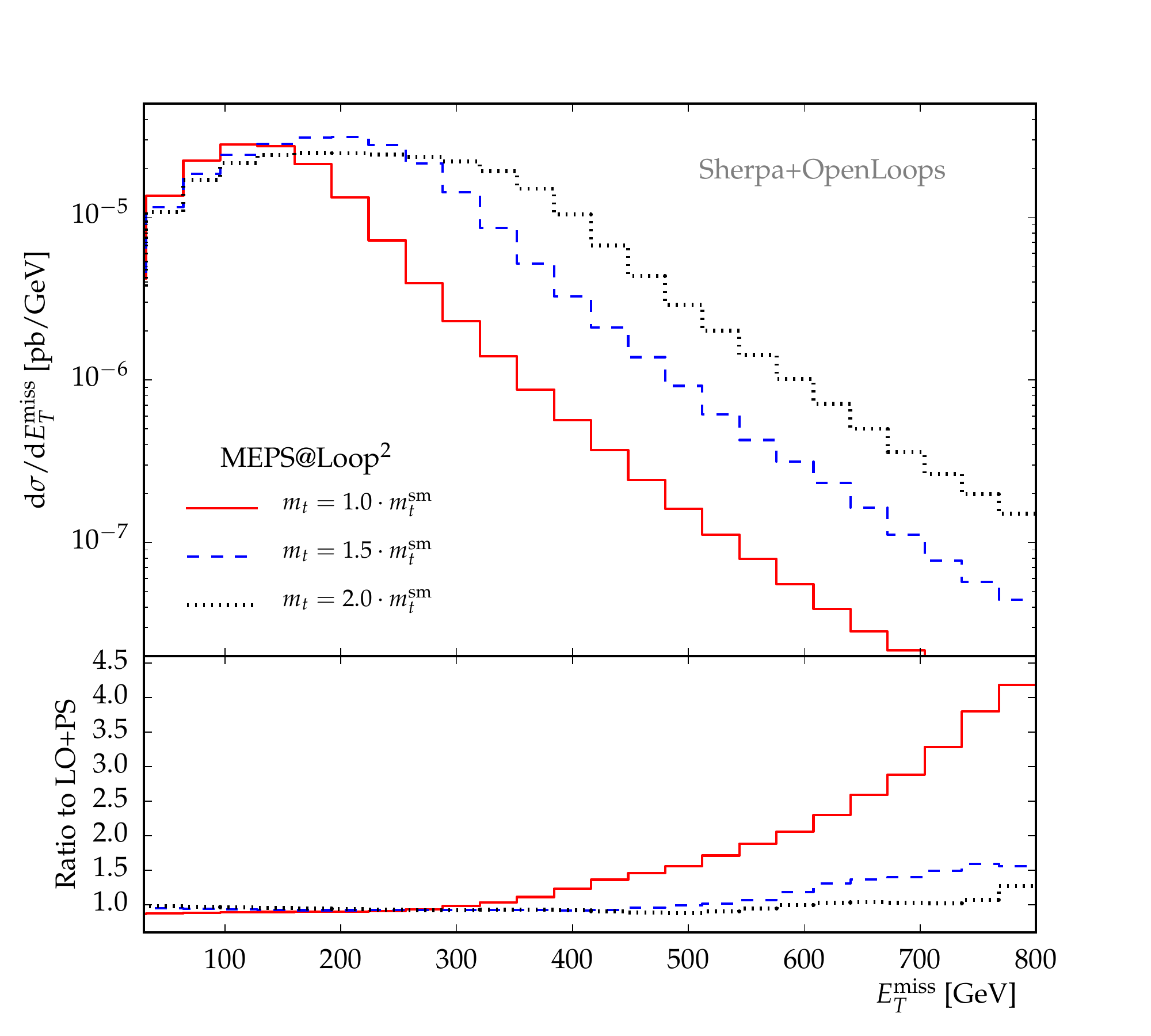}
  \includegraphics[width=.5\textwidth]{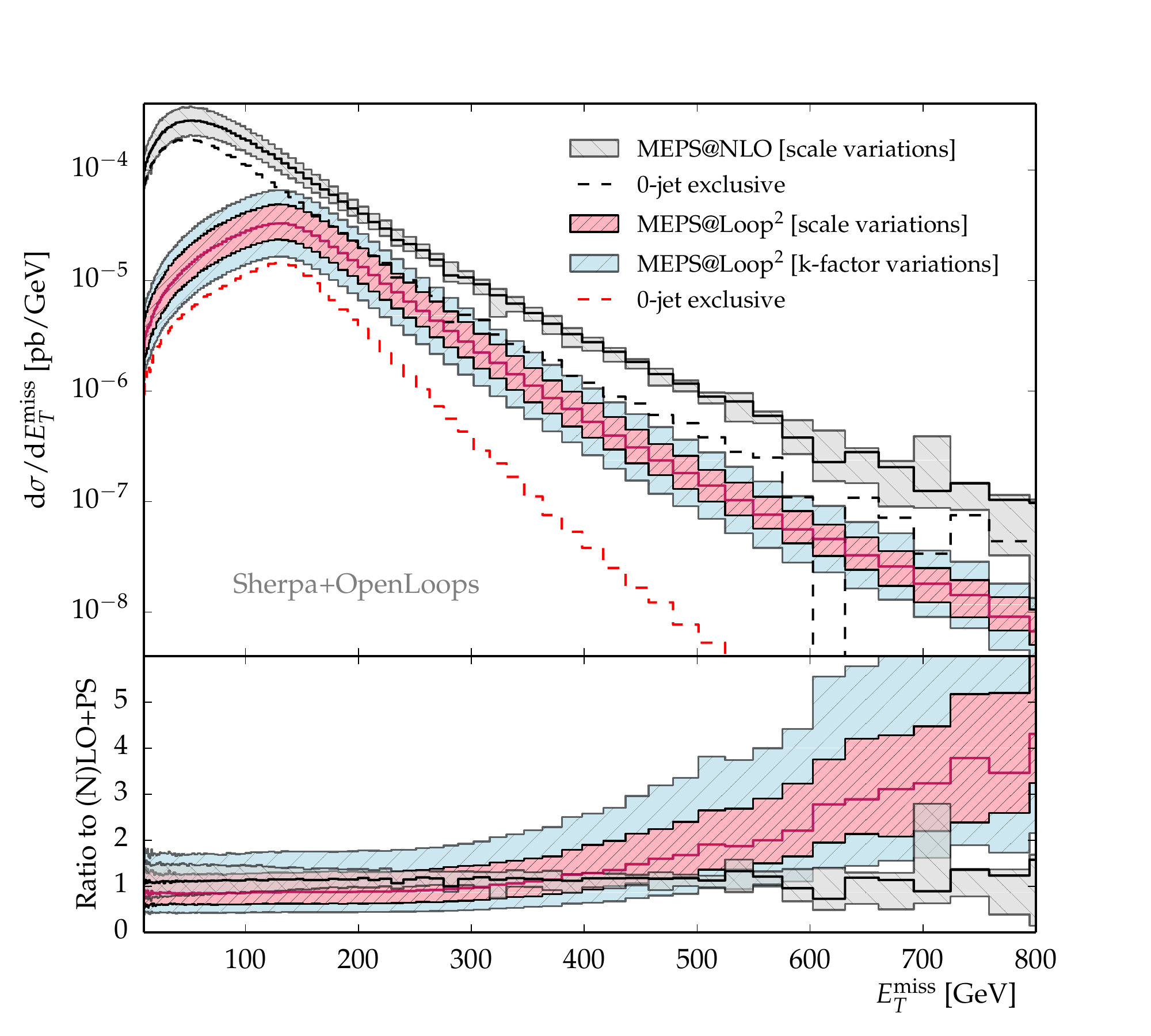}
  \caption{Missing transverse energy distributions after basic selection cuts
    in the signal channel of the invisible Higgs decay search, assuming a
    branching fraction for $H\to inv$ of $1$. Left panel: Varying the
    mass of the top-quark running in the loop in the GF contribution shows the
    threshold effect, extending to the tail of the distribution.  Right panel:
    Uncertainty bands obtained from scale variations along with $K$-factor
    variations for the DY and GF contributions. The NLO Drell-Yan and the
    loop-induced gluon fusion are both merged up to one jet. The
    bottom panel presents the ratio between the $\text{MEPS@Loop}^{2}$ to the
    $\text{Loop}^{2}$+PS and the $\text{MEPS@NLO}$ to the \textsc{MC@NLO}.}
 \label{fig:inv}
\end{figure}

We start the study of the signal sample by applying some typical basic
selection cuts.  We require two same-flavour opposite charged leptons with
transverse momentum $p_{Tl}>20~\gev$ in the pseudo-rapidity range $|\eta_l|<2.5$
and an invariant mass in the region $|m_{ll}-m_Z|<15~\gev$.  Jets are
reconstructed with the anti-$k_T$ algorithm with resolution parameter $R=0.4$
and $p_{Tj}>30~\gev$ in the pseudo-rapidity range $|y_j|<5$, using
\textsc{FastJet}~\cite{fastjet}.  After applying these kinematic selections, the
total signal cross section is strongly dominated by the DY component while
the GF mode contributes with only $\mathcal{O}(10\%)$ to the total $ZH$ cross
section~\cite{zh_gf}.  This finding seems to allow ignoring the GF channel for all
practical purposes.  However, selection cuts in searches for invisible Higgs
decays typically include a minimum $\met$ requirement, drastically reducing
the overwhelmingly large $t\bar{t}$+jets and $V$+jets backgrounds. As shown
in Fig.~\ref{fig:xs_minmet}, applying such a cut substantially changes the
relative composition of the signal cross section, and it enhances the relative
contribution of the gluon fusion production mode to up to $\mathcal{O}(30\%)$
of the DY mode.
The origin of this increase can be mainly traced
back to a harder transverse momentum spectrum triggered by a top quark
threshold enhancement around $m_{ZH}\sim 2m_t$.  This is supported by the
finding in the left panel of Fig.~\ref{fig:inv}, where we have varied the
mass of the heavy quark running in the loop.  In contrast, the DY
contributions do not feature such an enhancement but rather show the typical
$s$-channel suppression for large energies.  Therefore, despite its small
contribution to the inclusive cross-section, the GF mode can become a
significant player in the boosted regime and a proper modelling of this
component is of vital importance. In the right panel of Fig.~\ref{fig:inv}
we show the theory uncertainties stemming from scale and $K$-factor variations,
as detailed above, for the DY and the GF mode.  Clearly, the $K$-factor
variation leads to large effects, as the factor of two applied in both
directions directly translates into an uncertainty, which is about twice as
large as the effect of the standard scale variation in the GF mode.  This
size, about 30\% or so, is typical for a merged sample at LO, especially in
view of the fact that it is at least of order $\mathcal{O}(\alpha_S^2)$.  In
contrast, the scale uncertainty on the DY sample is much smaller, about
10-20\%.


In the lower panels of figure~\ref {fig:inv} (right panel), we also compare the missing
transverse energy spectrum of the gluon fusion component obtained from a
simple LO matrix element plus parton shower simulation (\textsc{Loop$^2+$PS})
with the one obtained from a merged calculation (\textsc{MEPS@Loop$^2$}),
taking into account matrix elements with up to one extra jet. In the boosted
regime above $\met\ge m_t$, the \textsc{Loop$^2+$PS} simulation significantly
undershoots the spectrum of the one using merging technology, with the
discrepancy reaching around 100\% around ${\met \sim 500~\gev}$ and further
increasing with energy.  This discrepancy has a origin similar to the finite
top--mass effects in $H$+jets production studied in~\cite{buschmann2},
namely that the extra jet emissions significantly impact the loop structure
of the matrix elements.  One can artificially suppress this effect by
increasing the value of the top quark mass, thereby pushing the relevant
scale for any loop structure effects to higher energies.  This is shown in
the left panel of figure~\ref{fig:inv}, where the discrepancy between the
\textsc{Loop$^2+$PS} and \textsc{MEPS@Loop$^2$} simulations becomes much
less severe.

Apparently, the effects induced by higher multiplicity jet--emission matrix
elements are significant and beyond the scope of conventional parton showers
alone.  They can be accounted for by applying matrix element merging techniques
as demonstrated here, since they correctly fill those phase space regions
that are typically problematic for the parton shower.  This provides a very
robust handle on theory uncertainties related to the application of vetoes
in searches for invisibly decaying Higgs bosons in $ZH$ production, which
quite often are an important feature in the search strategies.  Such vetoes on
extra jet emissions are commonly used, cf.\ for example~\cite{zh_atlas}, to
suppress the backgrounds associated with Higgs-Strahlung, such as top-pair
production and similar.  Following our discussion until now, we anticipate
that a jet veto will further suppress the fraction of the loop-induced signal component, 
even  when large $\met$ is being required.  A nice way to
have some idea about the impact of a jet veto is  to remind ourselves that
the no-emission probability of an additional parton or jet with a
transverse momentum $p_\perp$ off a quark $q$ or a gluon $g$ can be roughly
estimated, to leading logarithmic precision, using Sudakov form factors.
Schematically they are given by
\begin{equation}
  \Delta_{q,g}(\mu_Q,\,p_\perp)\,=\,\exp\left[
    -C_{F,A}\int\limits_{p_\perp^2}^{\mu_Q^2}\,\frac{\mathrm{d}q_\perp^2}{q_\perp^2}\,
    \frac{\alpha_S(q_\perp^2)}{\pi}\,
    \left(\log\frac{\mu_Q^2}{q^2}-\gamma_{q,g}\right)
    \right]\,,
\end{equation}
where $C_F=4/3$ and $C_A=3$ are the colour charges of the quark and gluon and
$\gamma_{q,g}$ are given by
\begin{equation}
  \gamma_q\,=\,-\frac{3}{2}\;\;\;\mbox{\rm and}\;\;\;
  \gamma_g\,=\,-\frac{\beta_0}{C_A}\,=\,-\frac{11}{6}+\frac{n_F}{9}\, ,
\end{equation}
with $n_F$ the number of active flavours.  The occurrence of the colour factors
easily motivates why the probability for not emitting a jet is larger
for quark than for gluon induced processes.

\begin{figure}[h!]
\includegraphics[width=.5\textwidth]{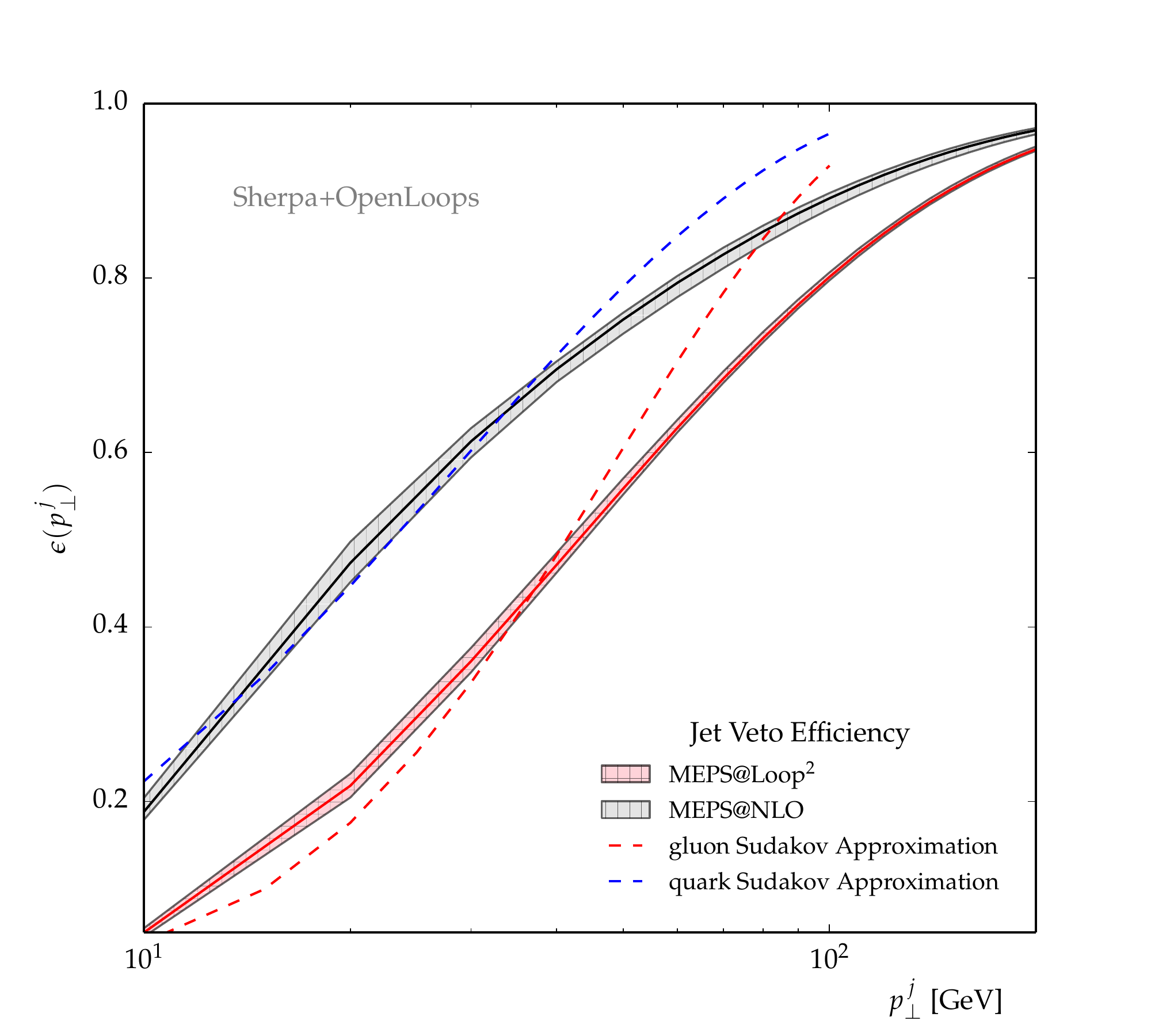}
\caption{Jet veto efficiencies for Drell-Yan-like contributions
  simulated via \textsc{MEPS@NLO} and for the gluon fusion component
  calculated via \textsc{MEPS@Loop$^2$}. We compare these predictions
  to simple Sudakov approximations.}
  \label{fig:jveto}
\end{figure}

Defining jet veto efficiencies as
\begin{align}
\epsilon(p^j_\perp) =
  \frac{\sigma_{0-\mathrm{jet}}^\mathrm{excl}(p^j_\perp)}{\sigma^\mathrm{incl}}\,,
\end{align}
that is the fraction of the inclusive cross section which survives a
jet veto applied to jets above a certain transverse momentum cut $p^j_\perp$,
we confront in Fig.~\ref{fig:jveto} the simple Sudakov approximation for
jet vetoes in the production of colour singlet systems, {\it i.e.}\ $ZH$ final
states with an invariant mass of $m=m_H+m_Z$, with the exact
results stemming from our more detailed simulation.  It is remarkable in
how far the simple approximation is able to reproduce the more exact result
in the limit of small transverse momentum cuts applied in the jet veto.
The results shown in the figure confirm that in the experimentally relevant
ranges around \SI{20}{\giga\electronvolt}, the gluon fusion contribution is
largely suppressed by jet vetoes due the initial state gluon's propensity to
radiate.

\begin{figure}[hb!]
\includegraphics[width=.49\textwidth]{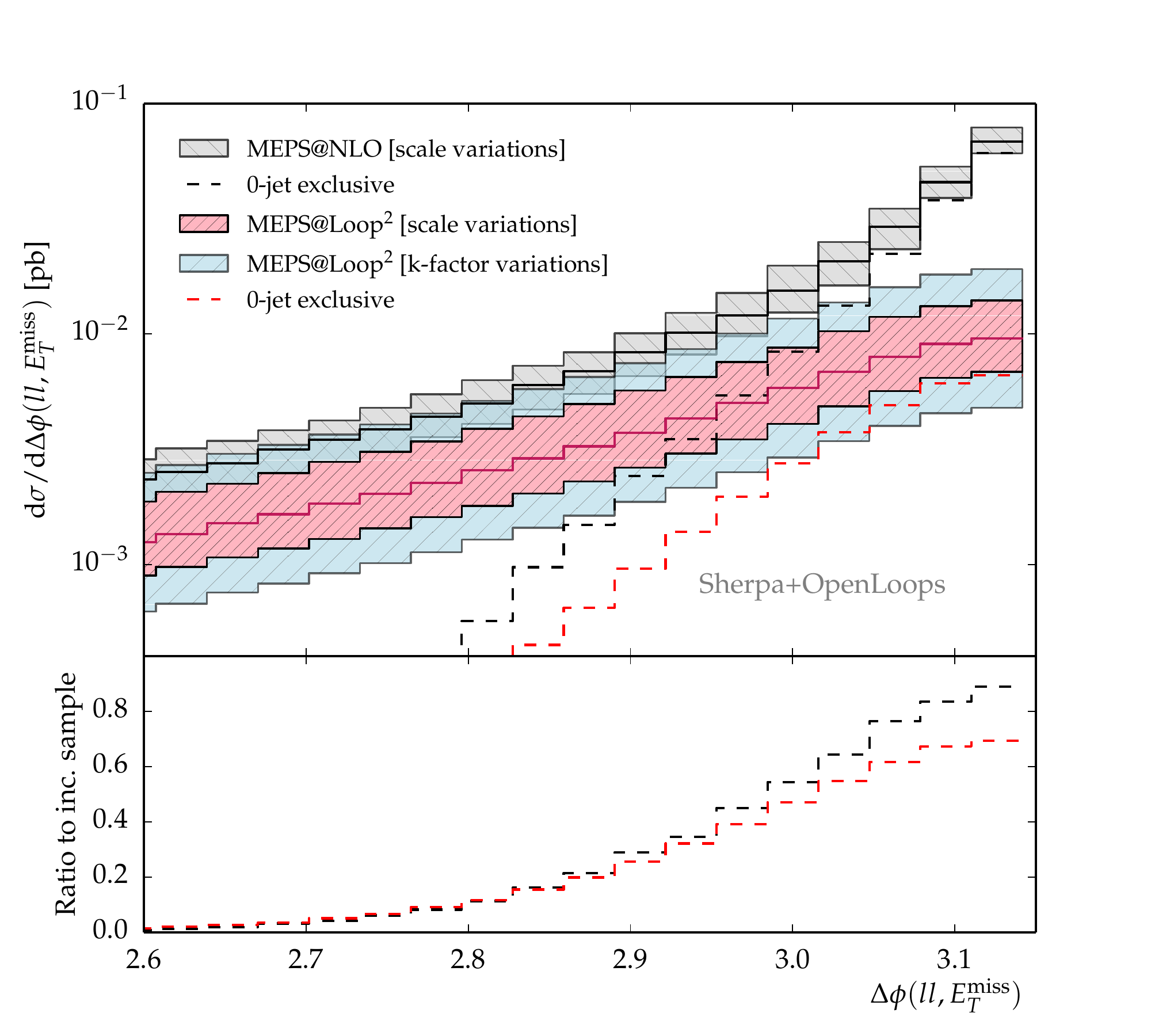}
\includegraphics[width=.49\textwidth]{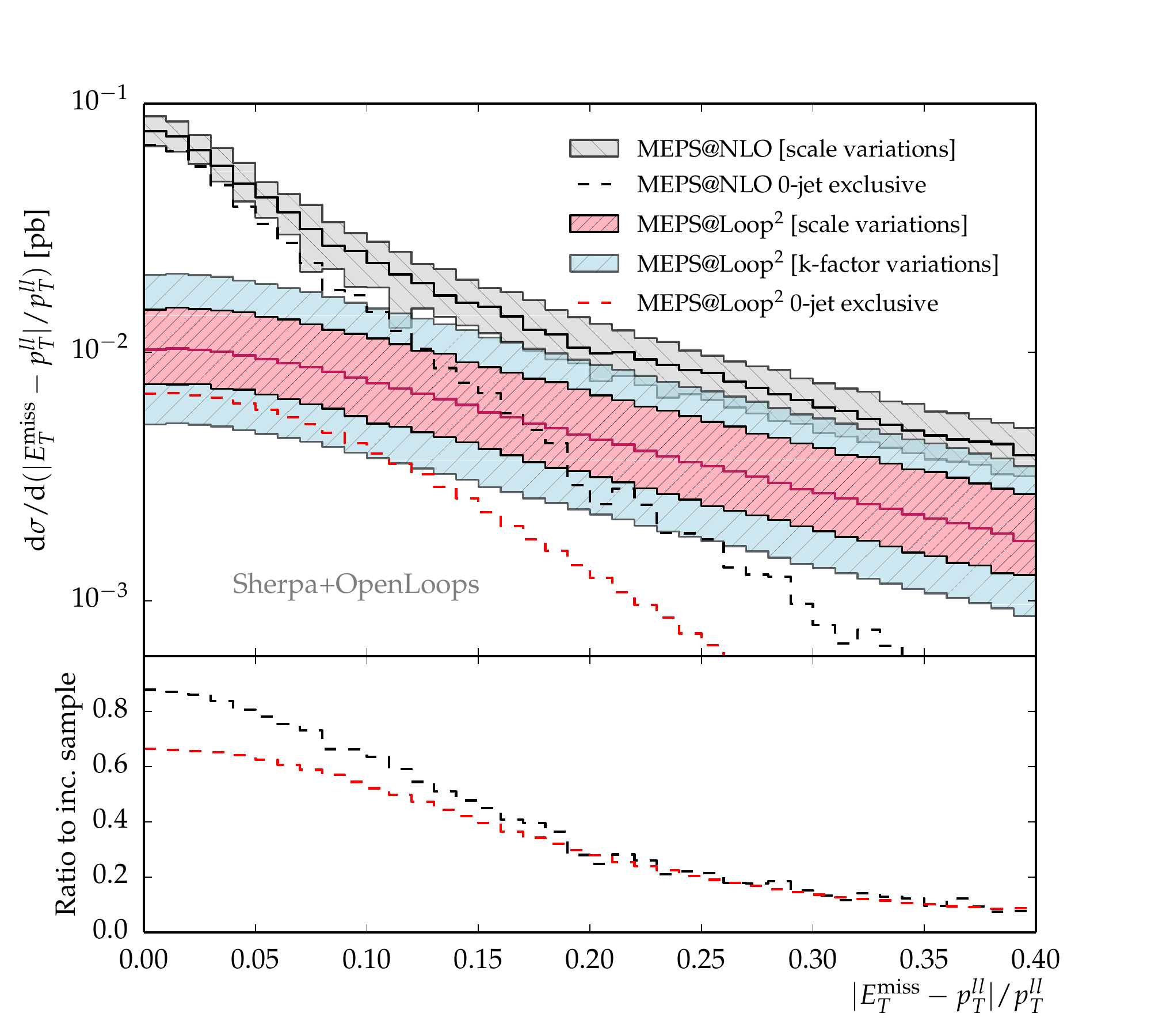}
\caption{Differential distributions for the major kinematic background
suppression cuts in invisible Higgs decay searches.}
\label{fig:zhcorr}
\end{figure}

A somewhat alternative way to suppress backgrounds is based on the observation
that at large $\met$, the $Z$ and $H$ bosons tend to be more or less back-to-back, 
rendering a selection cut on their relative azimuthal angle $\Delta\phi(ll,\met)$ an
efficient means to improve the signal-to-background ratio.  This is because additional
jets would decorrelate the $Z$ and the $H$ in the transverse plane will
effectively be vetoed by such a cut.  As shown in Fig.~\ref{fig:zhcorr},
these assumptions hold for the Drell-Yan-like contributions. The distributions
peak strongly at large azimuthal separations $\Delta\phi(ll,E_T^\mathrm{miss})$
and at small values of $|E_T^\mathrm{miss}-p_T^{ll}|/p_T^{ll}$.  In the case of
the gluon fusion contribution, however, the enhancement of the distributions
is much less pronounced in these regions, due to the larger level of QCD
radiation decorrelating the $ZH$ pair; configurations in which the Higgs
recoils against a jet rather than the $Z$ have a strong impact here.
This is even the case at $|E_T^\mathrm{miss}-p_T^{ll}|/p_T^{ll}=0$ and
$\Delta\phi(ll,E_T^\mathrm{miss})=2\pi$.  As can be seen in the lower panels of
Fig.~\ref{fig:zhcorr}, even in this kinematic regime there is a significant
contribution from one-jet events in case of the gluon fusion component,
whereas this region is depleted of 1-jet events for DY topologies.
Performing a search binned in jet multiplicities, as it was done in reference
\cite{zh_cms}, therefore retains sensitivity to the gluon fusion component.
In fact, as shown in Tab.~\ref{tab:cuts_inv}, the gluon fusion component
can be as large as 40\% after applying typical selection cuts in
the 1-jet inclusive bin. Modelling this very contribution reliably requires
the additional jet emission matrix elements and makes the merging techniques
applied here an indispensable tool.

\setlength{\extrarowheight}{2pt}
\begin{table}[h!]
\begin{tabular}{ >{$}l<{$\hspace{3mm}} >{$}c<{$} >{$}c<{$} >{$}c<{$\hspace{3mm}} >{$}c<{$} >{$}c<{$} >{$}c<{$}}
  & \multicolumn{3}{c}{MEPS@NLO}
  & \multicolumn{3}{c}{MEPS@Loop$^2$} \\
  & \sigma_\text{incl}[\si{\fb}]
  &  \sigma^{0-\text{jet}}_\text{excl}[\si{\fb}]
  &  \sigma^{1-\text{jet}}_\text{incl}[\si{\fb}]
  & \sigma^\text{incl}[\si{\fb}]
  &  \sigma^{0-\text{jet}}_\text{excl}[\si{\fb}]
  &  \sigma^{1-\text{jet}}_\text{incl}[\si{\fb}] \\
  \hline
  |m_{ll}-m_Z|<\SI{15}{\giga\electronvolt},
  p_{Tl}>\SI{20}{\giga\electronvolt}, |y_l|<2.5 & 34.5^{+9.1}_{-7.7} & 21.1^{+5.3}_{-4.5} & 13.4^{+4.1}_{-3.2} & 4.9^{+2.4}_{-1.4} & 1.74^{+0.8}_{-0.51} & 3.2^{+1.6}_{-0.9} \\
  \met>\SI{120}{\giga\electronvolt} & 9.7^{+1.8}_{-1.5} & 4.98^{+0.88}_{-0.69} & 4.74^{+0.95}_{-0.82} & 2.9^{+1.4}_{-0.8} & 0.95^{+0.45}_{-0.28} & 1.96^{+0.97}_{-0.56} \\
  \Delta\phi(ll,E_T^\mathrm{miss})>2.5 & 8.0^{+1.5}_{-1.3} & 4.97^{+0.88}_{-0.69} & 3.04^{+0.61}_{-0.57} & 2.4^{+1.2}_{-0.7} & 0.95^{+0.45}_{-0.28} & 1.42^{+0.74}_{-0.41} \\
  |p_T(ll)-E_T^\mathrm{miss}|/p_T(ll)<0.25 & 6.5^{+1.2}_{-1} & 4.81^{+0.83}_{-0.65} & 1.65^{+0.33}_{-0.32} &1.57^{+0.78}_{-0.46} & 0.88^{+0.41}_{-0.26} & 0.70^{+0.37}_{-0.21}
\end{tabular}
\caption{Cut flow for typical selection cuts in invisible Higgs decay
  searches. We list the individual contributions from Drell-Yan
  production modes \textsc{MEPS@NLO} and the loop-induced
  \textsc{MEPS@LOOP$^2$} component in ${pp\rightarrow (H\rightarrow
    \textit{inv})(Z\rightarrow e^+e^-,\mu^+\mu^-)}$ at the LHC $\sqrt{s}=13~\tev$.
  Uncertainties are obtained from scale variations as described in the
  text. For the loop induced contributions, they become as large as the ones one would obtain from varying
  the K-Factor in some cases. This is despite the fact that the $K$-factor variation
  error bands of the differential distributions in Figs.~\ref{fig:inv}
  and~\ref{fig:zhcorr} exceed the scale variation error bands considerably.}
\label{tab:cuts_inv}
\end{table}

Apparently however, the gluon fusion component was accounted for
neither by ATLAS nor by CMS in their searches for invisible Higgs
boson decays at Run I~\cite{zh_atlas, zh_cms}. We find that for the ATLAS
analysis, the gluon fusion component can indeed be neglected. A
jet-veto requirement in conjunction with selection cuts similar to the
ones shown in table \ref{tab:cuts_inv} suppress the GF component to
merely $\mathcal{O}$(\si{4}{\percent}) of the total signal rate. The
CMS analysis, however, takes the 1-jet exclusive bin into
account separately, thereby retaining sensitivity to the gluon fusion
component. We explicitly checked that the significance of the GF
contributions in the 1-jet exclusive bin at Run I energies is comparable
to our findings in table \ref{tab:cuts_inv}.

\subsection{Hadronic decays: $\boldsymbol{Z(ll)H(b\bar{b})}$}

We now analyse the Higgs-Strahlung channel for the $H\rightarrow b\bar{b}$
decay mode. In this case, the Higgs candidate is part of a multi-jet system that should
contain not only its decay products, but also the associated final state QCD
radiation.  This simple picture is blurred by initial state QCD radiation
and additional particles originating from the underlying event ``splashing''
into the fat-jet system stemming from the Higgs boson decay.  This complicated
final state renders proper modelling of the QCD emissions an indispensable
requirement for a successful and robust analysis of this process.  In this
section, we will therefore discuss, in particular, the relevance of multi-jet
merging techniques. \bigskip
\begin{figure}[ht!]
  \includegraphics[width=.49\textwidth]{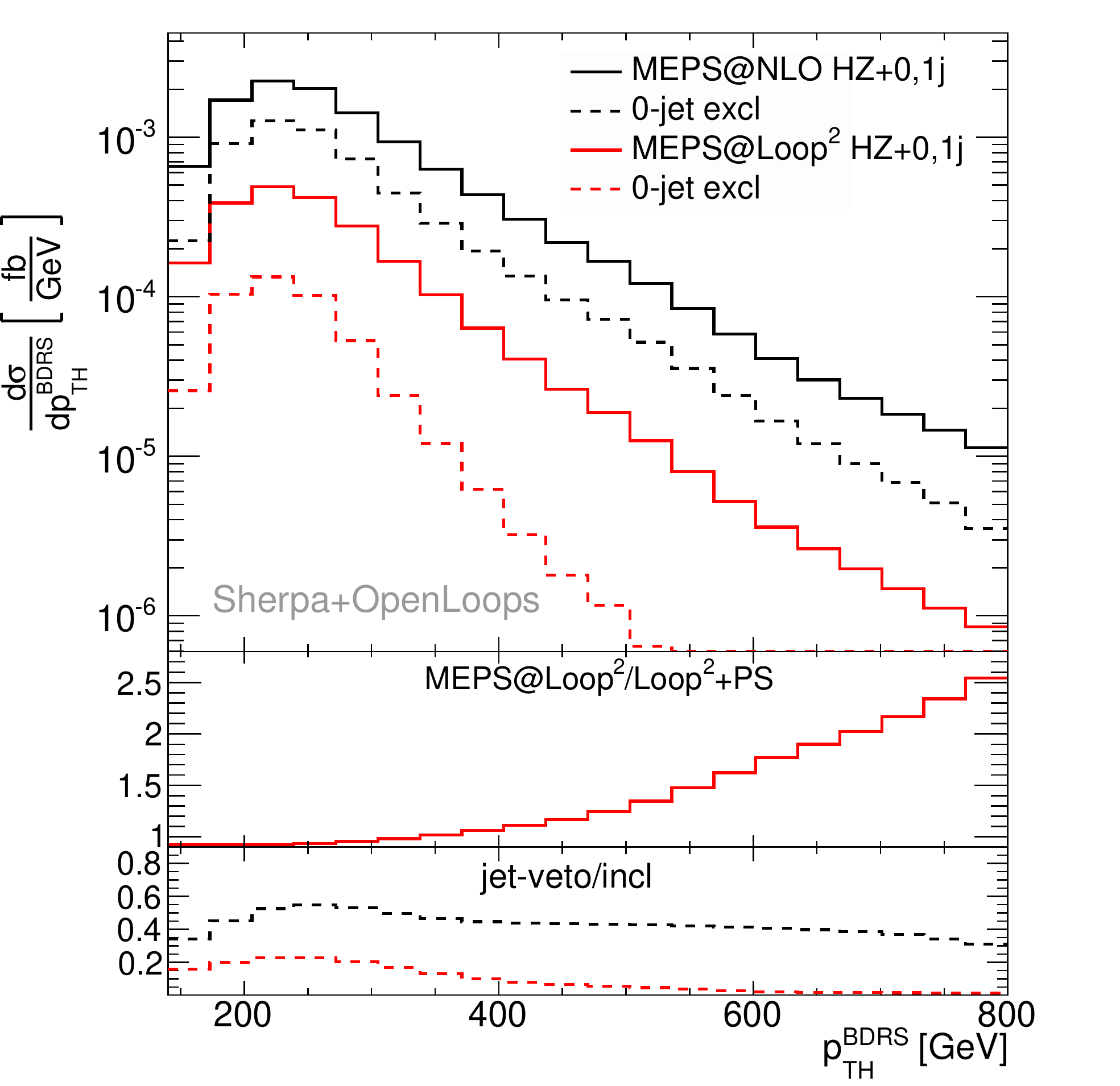}
  \includegraphics[width=.49\textwidth]{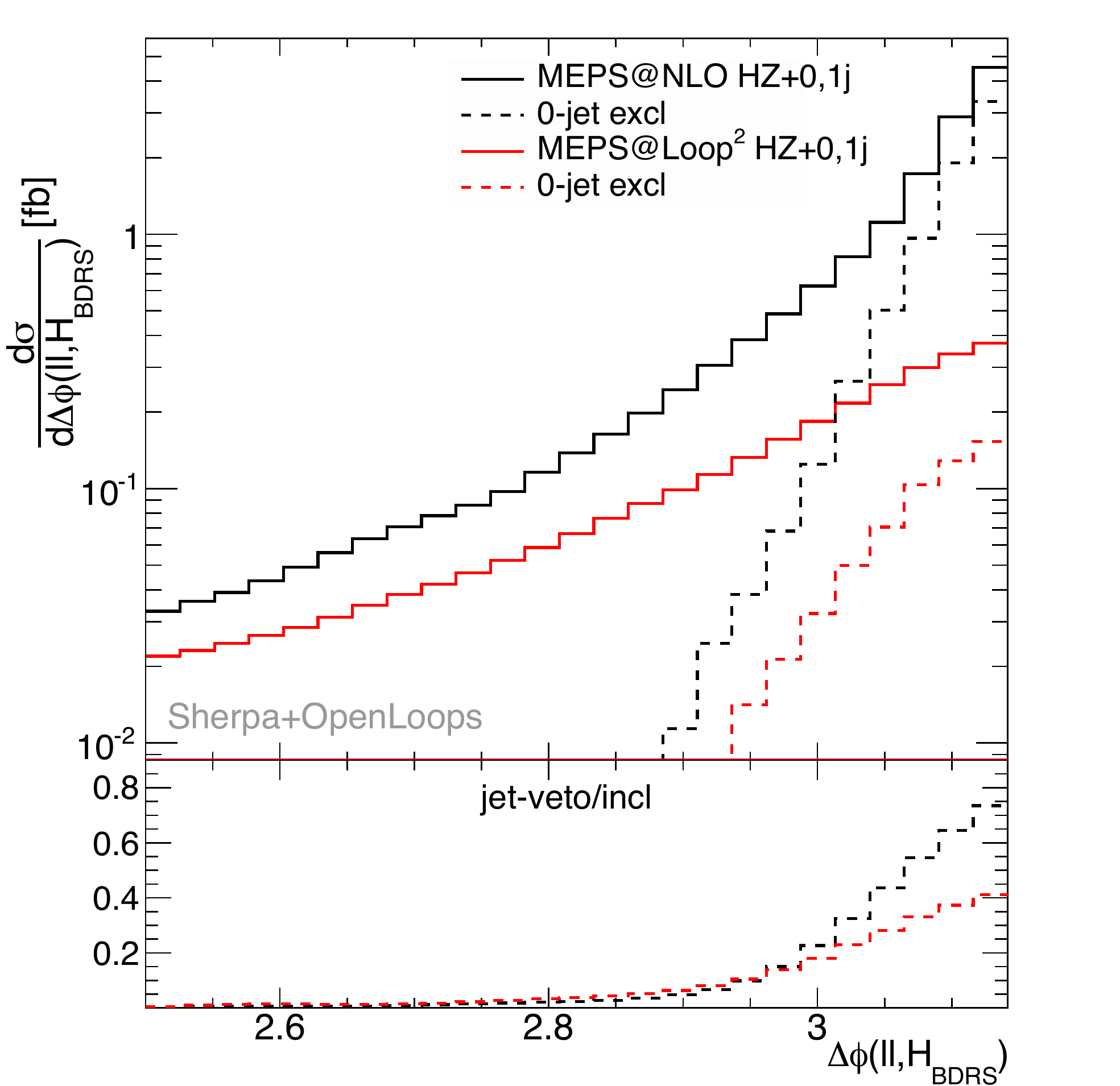}
  \caption{Transverse momentum distribution of the filtered Higgs candidate
    $p_{TH}^{\bdrs}$ (left) and azimuthal angle $\Delta\phi(ll,H_{\bdrs})$ (right)
    between the Higgs boson and the $Z$ candidate for ${pp\rightarrow (H\rightarrow
    b\bar{b})(Z\rightarrow e^+e^-,\mu^+\mu^-)}$ production at $\sqrt{s}=13~\tev$.
    The NLO-merged Drell-Yan contribution is shown in black and the loop-induced
    LO-merged gluon fusion mode in red. The bottom panel presents the ratio between
    the jet-vetoed and the inclusive sample and the central panel (left only)
    presents the ratio between $\text{MEPS@Loop}^{2}$ merged up to one jet and the
    parton shower sample.}
 \label{fig:hbb}
\end{figure}

To highlight the effect of higher order effects and to quantify the impact
of multi-jet merging, we follow the BDRS analysis~\cite{bdrs} as a well
understood benchmark:  First, we impose some basic selection cuts requiring
two same-flavour opposite charged leptons with transverse momentum
$p_{Tl}>30~\gev$, pseudo-rapidity $|\eta_l|<2.5$ and invariant mass in the
window ${75~\gev<m_{ll}<105~\gev}$.  We then impose the reconstructed $Z$
boson to have a large transverse momentum ${p_{T,ll}>200~\gev}$.  In the BDRS
algorithm the hadronic final states of events are clustered into fat jets
using the Cambridge-Aachen algorithm with radius $R=1.2$.  The analysis
demands at least one fat jet with $p_{TJ}>200~\gev$ and ${|\eta_J|<2.5}$,
acting as the candidate for the Higgs boson.  This candidate is tagged
through jet substructure techniques including the mass drop criteria and the
requirement of three filtered subjets for which the two hardest ones need
to be $b$-tagged.  Our analysis assume a flat 70\% $b$-tagging
efficiency and a 1\% mistag rate.

In Fig.~\ref{fig:hbb} (left panel) we display the filtered Higgs jet transverse momentum
$p_{TH}^{\bdrs}$.  In analogy to the $\met$ in the case of invisible decays above,
\textsc{MEPS@Loop$^2$} presents an enhancement with respect to the $ZH$
\textsc{Loop$^2$+PS}. This effect is noticeably smaller than in the invisible
scenario, however it is still relevant.   In the invisible search the
requirement of ${\met>120~\gev}$ leads to smaller invariant masses for the
combined $ZH$ system, which in turn set the scale for the parton shower to
populate the phase space with the emission of extra jets, while here the
cut on the $Z$ boson transverse momentum is at $200~\gev$.  The emission
phase space offered to the parton shower thus is larger in the
$H\to b\bar{b}$ case.  As a consequence, in $H\to inv$, the exact matrix
elements, only included through the merging, have a larger impact.
Additionally to this kinematic effect, the hadronic Higgs decay is naturally
more sensitive towards QCD radiation which also induces some
differences~\cite{banfi}.

\begin{figure}[b!]
  \includegraphics[width=.49\textwidth]{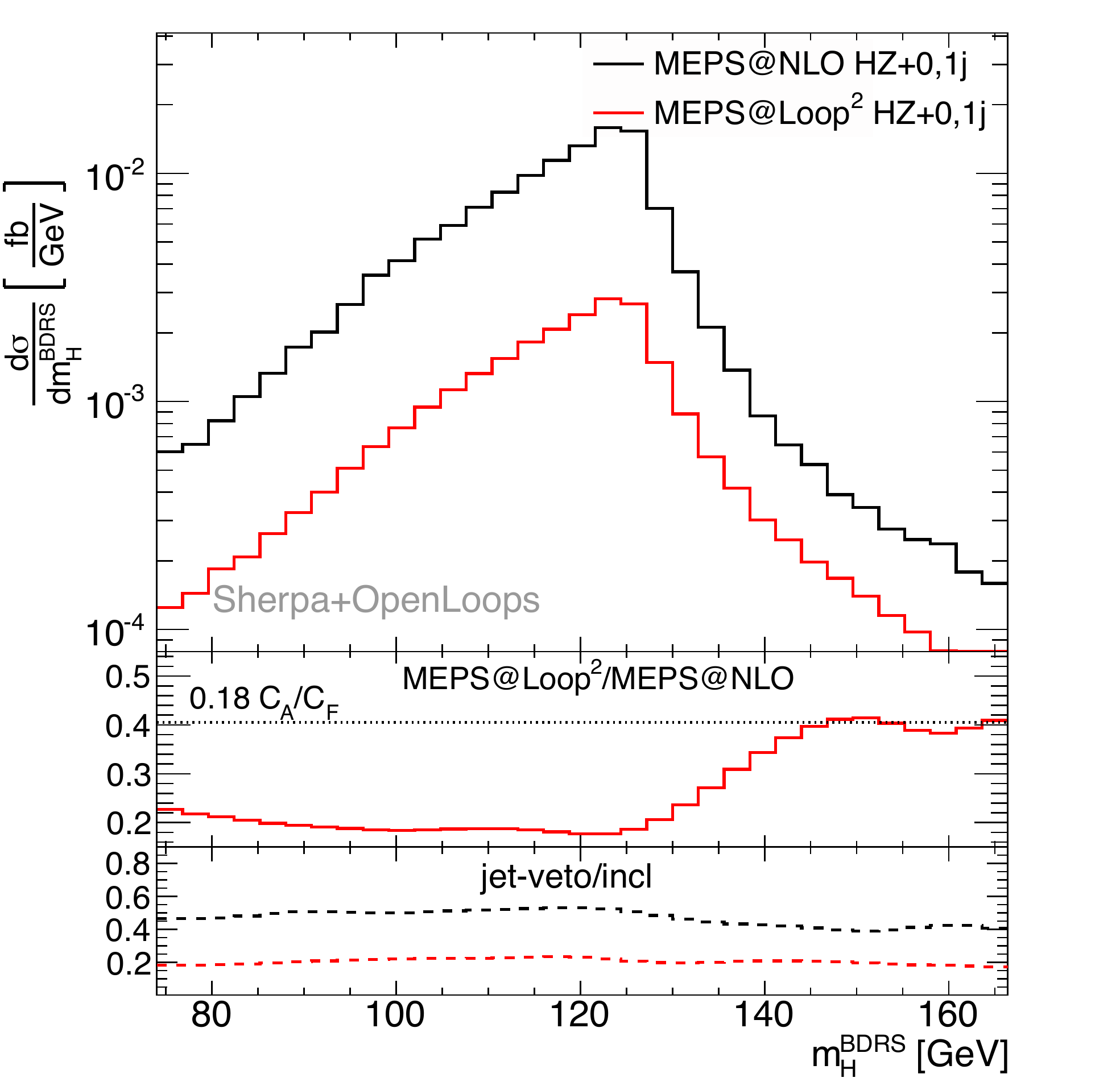}
  \caption{Invariant mass distribution of the filtered Higgs fat jet
    candidate $m_{H}^{\bdrs}$ in
    ${pp\rightarrow (H\rightarrow b\bar{b})(Z\rightarrow e^+e^-,\mu^+\mu^-)}$
    at $\sqrt{s}=13~\tev$.  The NLO DY contribution is shown in black and
    the loop-induced GF mode in red. Both samples are merged up to one jet.
    The central panel presents the ratio between the $\text{MEPS@Loop}^{2}$
    to the \textsc{MEPS@NLO} and the bottom panel the ratio between the
    jet-veto to the inclusive samples.}
 \label{fig:mbb}
\end{figure}

In Fig.~\ref{fig:mbb} we present some of the QCD radiation dynamics in the
invariant mass distribution for the filtered Higgs.  We first notice that the
corrections to the production of the Higgs boson, that are performed here via
the \textsc{MEPS@NLO} and \textsc{MEPS@Loop$^2$}, contribute to the high mass
tail $m_{H}^{\bdrs}\ge 125~\gev$. Extra parton emissions from the fixed order or
from the parton shower arising from the $ZH$ production can be reclustered in
the Higgs fat-jet, therefore enhancing its mass. On the other hand, the
invariant mass region $m_{H}^{\bdrs}\le 125~\gev$ is populated from the shower
radiation off the Higgs decays,  therefore decreasing the reconstructed
invariant mass $m_{H}^{\bdrs}$. Based on these observations we can for instance
understand the profile of the gluon fusion contribution with respect to the
Drell-Yan-like shown in the bottom panel.  While their effects at
$m_{H}^{\bdrs}\le 125~\gev$ have the same source (\ie shower emissions off the
$b\bar{b}$ pair), at $m_{H}^{\bdrs}\ge 125~\gev$  we observe  a big enhancement
that goes from $\mathcal{O}(18\%)$ to $\mathcal{O}(40\%)$ of the DY rate. This
is a side effect of the  larger radiation pattern arising from the gluon fusion
component, which benefits in particular from the larger initial state colour
charge $C_A/C_F=9/4$. Indeed, that captures  the size of the enhancement in a
very good approximation $0.18\times 9/4\sim0.4$. Besides, we notice that the jet
veto suppresses the cross-section by approximately a constant factor over the
full $m_{H}^{\bdrs}$ distribution. This clearly indicates that both the parton
shower off the Higgs decays and the  corrections to the production are properly
covering all the important phase space regions.\bigskip

The $\Delta\phi(ll,H_{\bdrs})$ distribution shows that typical selection cuts of
order $\Delta\phi(ll,H_{\bdrs})>2.5$ have a subleading impact on the inclusive
rates for the signal component, see Fig.~\ref{fig:hbb} (right panel).
Additionally, in the presence of an extra jet veto this azimuthal correlation
requirement can be pushed even further, almost  without extra losses. In
Table~\ref{tab:cuts_hbb} we display the impact of the cut
$|m_{H}^{\bdrs}-m_{H}|<10~\gev$ and the extra-jet veto requirement. While the
cut $\Delta\phi(ll,H_{\bdrs})>2.5$ has an imperceptible impact to both
components, the extra-jet veto weakens the gluon-fusion signal contribution to a
subleading level. Thus, if possible  one should use other handles (than
extra-jet vetoes) to the background suppression, especially for possible BSM
studies intrinsically  associated to the loop-induced component.  We will
further comment on this in the following section.

\begin{table}[h!]
\begin{tabular}{l || c  | c || c | c  }
  \multicolumn{1}{c||}{} &
  \multicolumn{2}{c||}{MEPS@NLO}&
  \multicolumn{2}{c}{MEPS@Loop$^2$}
   \\
  \hline
 \multirow{1}{*}{cuts} &
 \multicolumn{1}{c|}{$\sigma_{incl}$}  &
 \multicolumn{1}{c||}{$\sigma_{0-jet}$}&
 \multicolumn{1}{c|}{$\sigma_{incl}$}  &
 \multicolumn{1}{c}{$\sigma_{0-jet}$} \\
  \hline
BDRS reconstruction
 & 0.37  & 0.18 & 0.07 & 0.02 \\
$|m_{H}^{\bdrs}-m_H|<10~\gev$
 & 0.16 &  0.09 & 0.03 & 0.01 \\
\end{tabular}
  \caption{Cross sections for the Drell-Yan  and loop-induced components of
    ${pp\rightarrow (H\rightarrow b\bar{b})(Z\rightarrow e^+e^-,\mu^+\mu^-)}$
    production at LHC  $\sqrt{s}=13~\tev$. Both samples are merged up to one jet and
    the selection cuts follow the BDRS analysis that is described in the text. The
    rates are given in fb and account to 70\% b-tagging efficiency. Hadronisation
    and underlying event effects are accounted for. }
\label{tab:cuts_hbb}
\end{table}

\section{Higgs-Strahlung: Boosting Coupling Constraints}
\label{sec:yukawa}

\begin{figure}[t!]
  \includegraphics[width=.49\textwidth]{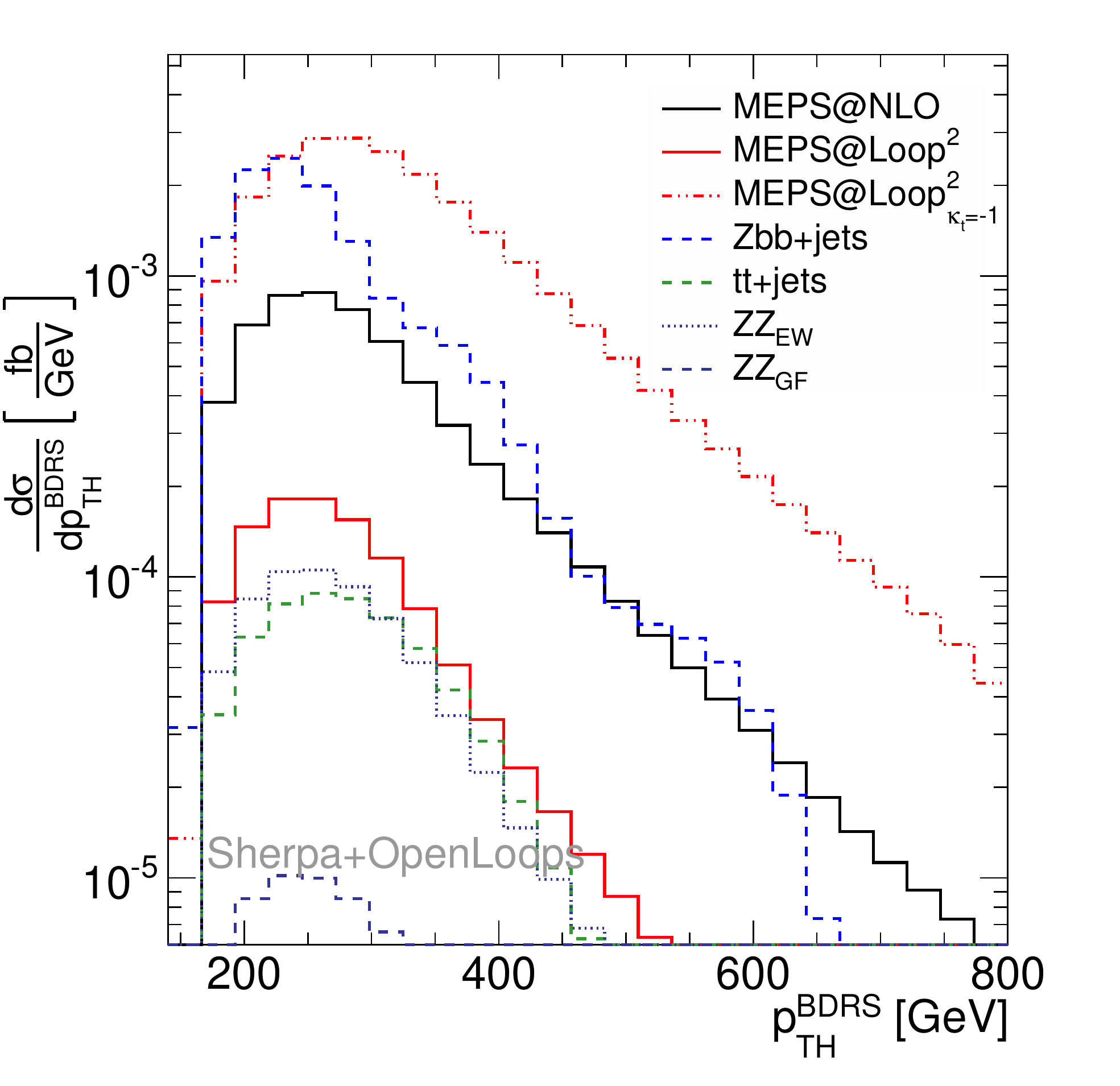}
 \caption{Transverse momentum distribution of the filtered Higgs candidate $p_{TH}^{\bdrs}$  for
the Higgs-Strahlung signal and backgrounds at $\sqrt{s}=13~\tev$. We also display the loop-induced
gluon fusion component for the BSM hypothesis with negative top Yukawa, $\kappa_t=-1$.}
 \label{fig:kappas}
\end{figure}

In this section we analyse the constraining power to the top-quark Yukawa coupling  that can be derived from the
${pp\rightarrow (H\rightarrow b\bar{b})(Z\rightarrow e^+e^-,\mu^+\mu^-)}$  production.  Following Eq.~\ref{eq:kappas},
we notice that the Drell-Yan component does not present any dependence on the top Yukawa $\kappa_t$ but only on the
size of the $HVV$ coupling  $\kappa_V$. On the other hand, the gluon fusion develops a dependence  on both $\kappa_V$ and
 $\kappa_t$\footnote{Only the relative sign between $\kappa_V$ and $\kappa_t$ is physical, thus only positive $\kappa_V$
is considered without loss of generality.}. Hence, we can write the matrix element for the $ZH$ production as
\begin{alignat}{5}
\mathcal{M}&
= \kappa_t \mathcal{M}_t +
\kappa_V \mathcal{M}_V
\; ,
\label{eq:amplitude}
\end{alignat}
where the $t$ stands for the top Yukawa contributions and $V$ for the contributions proportional to the  $HVV$ coefficient.
The dependence on these coefficients can be straightforwardly  translated to the Higgs $p_{TH}$ spectrum via
\begin{alignat}{5}
\frac{d\sigma}{dp_{TH}} &
= \kappa_t^2 \; \frac{d\sigma_{tt}}{dp_{TH}}
+ \kappa_t \kappa_V \; \frac{d\sigma_{tV}}{dp_{TH}}
+ \kappa_V^2 \; \frac{d\sigma_{VV}}{dp_{TH}}
\; .
\label{eq:distribution_boost}
\end{alignat}
Therefore, the $p_{TH}$ distribution encodes the information about both the size
and sign of the top Yukawa. To estimate the LHC sensitivity towards these
coefficients, we consider the major backgrounds for Higgs-Strahlung, namely
$t\bar{t}$+jets, $Zb\bar{b}$+jets, and $ZZ_{EW}$. Besides these standard
contributions, we also accounted for the loop-induced gluon fusion $ZZ_{GF}$
production depicted Fig.~\ref{fig:feynman_diagrams_interf}.  The
interferences with the QCD continuum as described in section~\ref{sec:mm}, see
Fig.~\ref{fig:feynman_diagrams_interf}, were shown to be subleading in this
analysis.
\begin{figure}[bh!]
 \includegraphics[width=.48\textwidth]{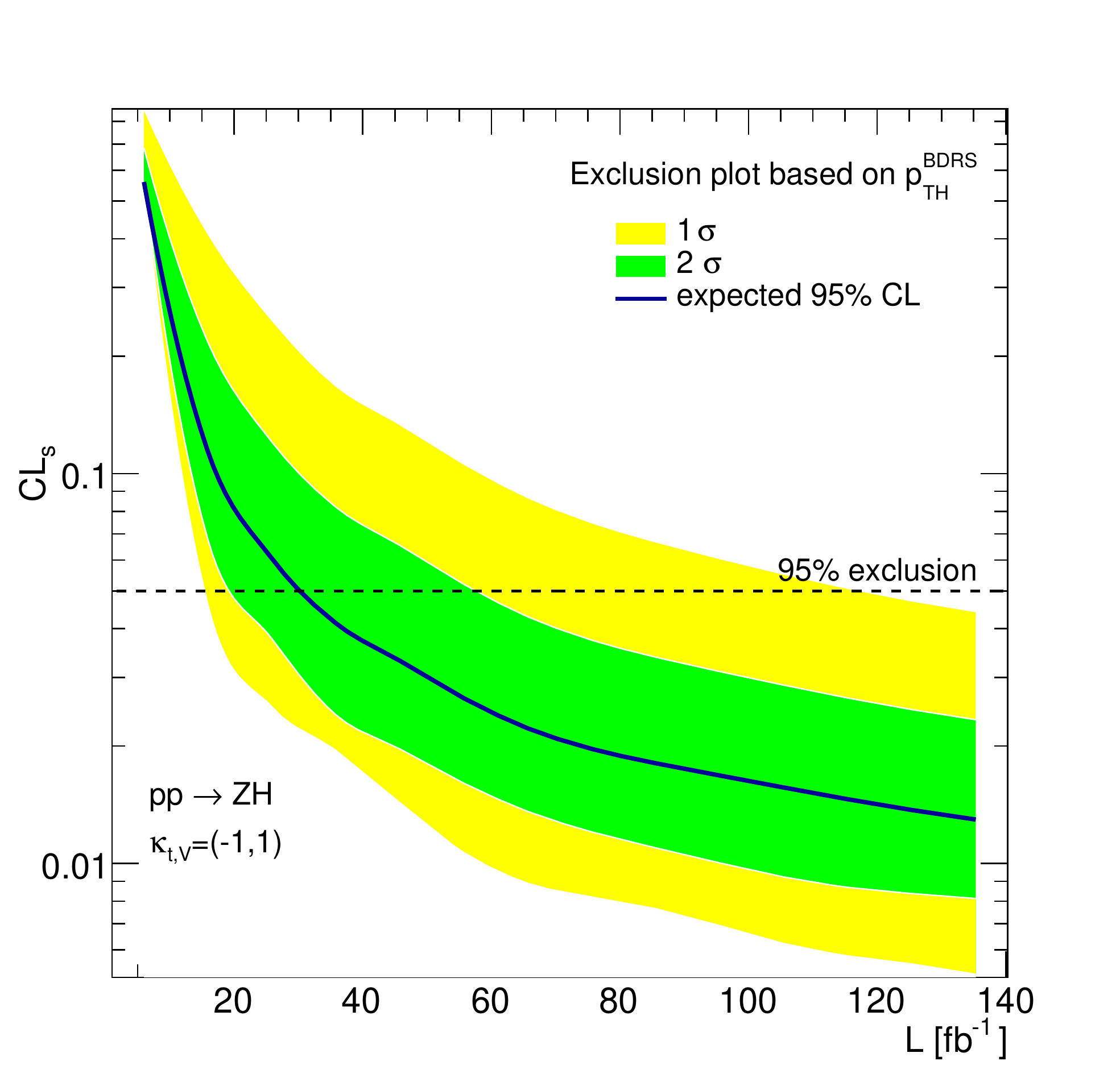}
 \caption{Confidence level for disentangling the negative top Yukawa hypothesis
    $\kappa_t=-1$ from the Standard Model. We display the results for
    ${pp\rightarrow (H\rightarrow b\bar{b})(Z\rightarrow e^+e^-,\mu^+\mu^-)}$ based
    on the $p_{TH}^{\bdrs}$ distribution.}
 \label{fig:CLs}
\end{figure}

As for the signal, it is important to properly  model the QCD radiation for the
background components. The $t\bar{t}$ is generated with the $0$-jet bin at NLO
and merged up to 3 jets simulated at LO. The $Zb\bar{b}$ and $ZZ_{EW}$ are
generated via \textsc{MC@NLO} and the loop-induced $ZZ_{GF}$ at LO. The cut-flow
is presented in Tab.~\ref{tab:cuts_analysis}. We avoid applying extra jet veto
requirements, since it would  deplete the GF signal component as derived  in the
previous section. Notice that  the loop-induced $ZZ_{GF}$ production presents a
non-negligible rate after the BDRS reconstruction, however the Higgs mass window
selection $|m_{H}^{\bdrs}-m_H|<10~\gev$  yields a subleading size to it.
\begin{table}[h!]
\begin{tabular}{l  || c  | c | c | c | c | c | c }
 \multirow{1}{*}{cuts} &
 \multicolumn{1}{c|}{$ZH_{GF}$ $\kappa_t=-1$}  &
 \multicolumn{1}{c|}{$ZH_{GF}$}  &
 \multicolumn{1}{c|}{$ZH_{DY}$} &
 \multicolumn{1}{c|}{$t\bar{t}$+jets}  &
 \multicolumn{1}{c|}{$Zb\bar{b}$+jets} &
  \multicolumn{1}{c|}{$ZZ_{EW}$}  &
 \multicolumn{1}{c}{$ZZ_{GF}$}    \\
  \hline
BDRS reconstruction
& 1.48 & 0.07 & 0.37  & 0.29  &  13.83 &0.79  & 0.10 \\
$|m_{H}^{\bdrs}-m_H|<10~\gev$
& 0.63 & 0.03 & 0.16 &  0.02 & 0.35  & 0.02 & 0.002\\
\end{tabular}
  \caption{Cut flow for $ZH$+jets (gluon fusion and Drell-Yan-like components),
    $t\bar{t}$+jets, $Zb\bar{b}$+jets and $ZZ$+jets. Furthermore, we generate the EW
    and loop induced QCD components.  All simulations were performed with
    \textsc{Sherpa+OpenLoops}.  The rates are given in fb and account  to 70\%
    b-tagging efficiency.  Hadronisation and underlying event effects taken into
    account.}
\label{tab:cuts_analysis}
\end{table}

In Fig.~\ref{fig:kappas} we present the signal and background transverse
momentum distributions after the selection cuts depicted in
Table~\ref{tab:cuts_analysis}. The background components are under control
through the whole spectrum with the $S/B$ ratio increasing towards higher
energies. The negative top Yukawa $\kappa_t=-1$ displays an amount of events
that surpasses the Drell-Yan and background components.  It largely benefits
from the $\sigma_{tV}$ term  that in the SM represents a destructive
interference in the  whole $p_{TH}^{\bdrs}$ distribution with a bigger magnitude
 than the other two terms $\sigma_{tt}$ and $\sigma_{VV}$ separately.

In Fig.~\ref{fig:CLs} we show the projection of the reach in the $ZH$ analysis
for the coupling determination. We  analyse the information from the different
$p_{TH}^{\bdrs}$ bins via the $CL_s$ method~\cite{Junk:1999kv} and estimate the
integrated luminosity necessary to exclude the negative top Yukawa solution at
95\% CL. A conservative systematic uncertainty of 50\% is inferred to the  GF
channel. The BSM hypothesis can be excluded with $\sim 30~\text{fb}^{-1}$ of
data.

\section{Summary}
\label{sec:summary}

We have studied the Higgs-Strahlung process at the LHC merging the zero and one
jet multiplicities for the Drell-Yan and loop-induced gluon fusion via the
\textsc{MEPS@NLO} and \textsc{MEPS@Loop$^2$} algorithms, respectively. We have
shown that the {\it multi-jet merging} is a fundamental ingredient to  properly
model the gluon fusion component. The merging leads to significant contributions
with respect to LO+PS simulations. For instance, for typical
$H\rightarrow\textit{invisible}$ searches at $p_{TH}\sim500~\gev$ the correction
factor is of order $\mathcal{O}(2)$.

A proper modelling of extra QCD emissions becomes even more important for the
$H\rightarrow b\bar{b}$ decay, since the Higgs candidate is part of this
multi-jet system.  We scrutinised the signal contributions at the boosted
kinematics and showed that \textsc{MEPS@NLO} and \textsc{MEPS@Loop$^2$} provide
a good description for the relevant distributions. In particular, we
observed significant improvements to the transverse momentum $p_{TH}^\mathrm{BDRS}$ and
reconstructed mass $m_{H}^\mathrm{BDRS}$ for the Higgs candidate.\medskip

Higgs-Strahlung search strategies often rely on extra jet veto requirements
that, however,  challenge the stability of perturbative expansions. We show that
\textsc{MEPS@NLO} and \textsc{MEPS@Loop$^2$} considerably decrease the impact
of jet vetoes on the uncertainties in comparison to \textsc{MC@NLO} and
\textsc{Loop$^2$+PS}, respectively. Furthermore, a larger suppression to the
loop-induced component is observed. At the boosted regime, in particular, extra
jet vetoes can deplete this signal component to subleading levels.

Finally, we also estimate the constraining power to the top Yukawa coupling via
jet substructure techniques for ${pp\rightarrow (H\rightarrow
b\bar{b})(Z\rightarrow e^+e^-,\mu^+\mu^-)}$ production. We perform a full
analysis accounting for all the major background  components that include, for
instance, the EW and loop-induced QCD $ZZ$ production. We conclude that the
Higgs-Strahlung can be used to access both the size and sign of the top Yukawa
coupling. Including conservative systematic uncertainties, the Run II LHC can
exclude at 95\% CL the negative top Yukawa solution $\kappa_t=-1$ with only
$\sim 30~\text{fb}^{-1}$.


\end{document}